\documentclass[12pt]{article}
\usepackage{epsfig, amsmath, amssymb}
\usepackage{graphicx,psfrag}
\usepackage{xcolor}
\newcommand{\tcb}{\textcolor {blue}}

\usepackage{slashed,verbatim,subfigure}
\usepackage{appendix}
\usepackage[colorlinks=true,linktocpage=true,
linkcolor=blue,citecolor=blue]{hyperref}
\usepackage[T1]{fontenc}
\usepackage{slashed,verbatim,subfigure}
\usepackage[numbers,sort&compress]{natbib}
\usepackage{amsmath}
\usepackage{mathrsfs}
\usepackage{amsbsy}
\usepackage{amssymb}
\usepackage{appendix}
\usepackage{graphicx}
\usepackage[final]{pdfpages}
\usepackage{dcolumn}
\usepackage{bm}

\newcommand{\be}{\begin{equation}}
\newcommand{\ee}{\end{equation}}
\newcommand{\ben}{\begin{equation}}
\newcommand{\een}{\end{equation}}
\newcommand{\bea}{\begin{eqnarray}}
\newcommand{\eea}{\end{eqnarray}}
\newcommand{\bA}{\begin{array}}
\newcommand{\eA}{\end{array}}
\newcommand{\bc}{\begin{center}}
\newcommand{\ec}{\end{center}}

\begin{document}


\begin{titlepage}

%

\bc

\hfill 
\\         [25mm]

{\Large \bf Communicating multiverses in a holographic \\ \vspace{3mm} de Sitter braneworld} 
\vspace{16mm}

{\textbf{Gopal Yadav}\footnote{{\small email: gopalyadav@cmi.ac.in/gopal12896@gmail.com}}} \\
\vspace{3mm}
{\small \it Chennai Mathematical Institute, \\}
{\small \it H1 SIPCOT IT Park, Siruseri 603103, India\\}

\ec
\vspace{25mm}

\begin{abstract}
 In this paper, we construct the wedge holography for the de-Sitter space as a bulk theory. First, we discuss a more general mathematical construction of wedge holography in parallel with wedge holography construction for the AdS bulk, and then we construct the wedge holography in the extended static patch. In the first case, we prove that one can construct wedge holography for the de-Sitter bulk for the static patch as well as global coordinate metric on end-of-the-world branes. Extended static patch wedge holography is constructed by joining the two copies of double holography in DS/dS correspondence. We find that wedge holography with extended static patch metric leads to the emergence of communicating universes. Further, we propose a model which provides the theoretical evidence of communicating multiverses.
\end{abstract}
\vspace{15mm}

\hspace{1.25cm}\fbox{\begin{minipage}{30em}{{\it One day there may well be proof of multiple universes...and in that universe, Zayn is still in one direction.  \hspace{13mm} {\bf Stephen Hawking}}}
\end{minipage}}

\end{titlepage}

{\footnotesize
\begin{tableofcontents}    
\end{tableofcontents}
}


\vspace{-5mm}

\section{Introduction}
It has always been fascinating to question whether our universe is the only universe, whether there are other universes in nature, or whether many universes exist in a multiverse. Is our universe part of a multiverse? Some questions related to Multiverse have been explored in \cite{Linde:2015edk,Kaku,Tegmark:2003sr,Nomura:2012zb,Bousso:2011up}. There is also progress done in this direction within braneworld by the use of wedge holography \cite{Yadav:2023qfg,Aguilar-Gutierrez:2023zoi} where the bulk theory is anti de-Sitter spacetime which has negative cosmological constant. In this work, we explore this type of question from the perspective of wedge holography where the bulk is de-Sitter space with positive cosmological constant.

Holography has played an important role in understanding various phenomena in many branches of physics. Originally, it was started as a duality between ${\cal N}$=4 supersymmetric Yang-Mills theory and type IIB string theory on $AdS_5 \times S^5$ background \cite{AdS-CFT}. We are able to decode the information about boundary theory from the bulk theory using holography and vice-versa. There are many generalizations of holography, e.g., dS/CFT correspondence \cite{Strominger:2001pn,dS-CFT-1} which is a duality between de-Sitter space in the bulk and Euclidean CFT living at the future/past boundary of de-Sitter space. This is a co-dimension one holography. In this paper, we discuss co-dimension two holography in the context of de-Sitter space. There is also so-called DS/dS correspondence \cite{Alishahiha:2004md}(see also \cite{Alishahiha:2005dj}) which describes duality between $d$-dimensional de-Sitter space and two CFTs which are coupled to each other and ($d-1$)-dimensional de-Sitter gravity. Further, see \cite{Mia:2009wj,Dhuria:2013tca,Yadav:2020tyo} for the application of holography to thermal QCD from a top-down approach, \cite{Hartnoll:2008vx,Ryu:2010hc} for condensed matter physics, \cite{Das:2006dz,Awad:2007fj,Awad:2008jf} where holography has been used in cosmology, \cite{Balasubramanian:2014hda,AlBalushi:2020kso,Jalan:2023dmq} for the discussion on multi-boundary wormholes. There is also a new version of holography in the context of AdS/CFT where time is emergent dimension known as ``Cauchy slice holography'' \cite{Araujo-Regado:2022gvw}\footnote{There are many applications of holography, and 25 years of the same was celebrated at ICTP-SAIFR (\url{https://www.ictp-saifr.org/holography25/}).}.

As an application of holography, we are able to resolve the information paradox \cite{Hawking}. For this purpose, there are three proposals: island proposal \cite{AMMZ,Almheiri:2020cfm}, double holography \cite{Almheiri:2019psy,Takayanagi:2011zk,Fujita:2011fp}, and wedge holography \cite{WH-1,WH-2}. In double holography, we couple the black hole living on an end-of-the-world brane with an external CFT bath living at the conformal boundary of the AdS bulk. This external bath acts as a sink to collect the Hawking radiation. Using double holography, the Page curve \cite{Page} has been obtained by computing the areas of Hartman-Maldacena \cite{Hartman:2013qma} and island surfaces. Further, efforts have been made to collect the Hawking radiation in a gravitating bath, and the setup is obtained by introducing two Karch-Randall branes in which one brane acts as a black hole, whereas another brane acts as a bath. The setup with two gravitating branes is known as wedge holography \cite{WH-1,WH-2}.

Wedge holography is a co-dimension two holography, which can be understood as follows. First, we localize the $d$-dimensional bulk Einstein gravity on $(d-1)$-dimensional Karch-Randall branes, which is called as braneworld holography \cite{KR1,KR2} and then the gravity living on Karch-Randall branes is dual to the CFT living at the $(d-2)$-dimensional defect because of AdS/CFT correspondence \cite{AdS-CFT}. Hence, $d$-dimensional Einstein gravity living in the wedge formed by Karch-Randall branes is dual to $(d-2)$-dimensional defect CFT living at the corner of wedge. For the AdS bulk, wedge holography has been explored in detail in \cite{GB-3,Massless-Gravity,Bhattacharya:2023drv,Li:2023fly,Miao:2024orm,Cui:2023gtf,Miao:2023mui,Miao:2023unv,Geng:2023iqd,Aguilar-Gutierrez:2023tic,WH-ii,WH-iii,Geng} and for flat spacetime bulk, see \cite{FS-Holography}. An interesting application of wedge holography is that it describes multiverse \cite{Yadav:2023qfg,Aguilar-Gutierrez:2023zoi} but in the AdS bulk spacetime. The multiverse is obtained by $2n$ Karch-Randall branes, which are joined at the common defect. Since there is Einstein gravity on each branch, therefore, we obtain a Multiverse made up of $2n$ universes \cite{Yadav:2023qfg}.

Wedge holography has been successfully constructed in the AdS and flat spacetime bulk. The motivation of the paper is to construct the wedge holography for the de-Sitter bulk. We will see that it is possible to construct wedge holography even for the de-Sitter bulk. This setup provides interesting results. It gives a hint for dS$_d$/CFT$_{d-2}$ correspondence, the existence of communicating universes, multiverse, and communicating multiverses within braneworld. We discuss two ways to construct wedge holography for the de-Sitter bulk: the first one is similar to the wedge holographic realization for the AdS bulk, and the second one is by gluing two copies of double holography in the extended static patch. Parallel universes are obtained by gluing many copies of extended static patch wedge holography; however, we can have ``disconnected parallel universes''. Whenever we join all the branes in disconnected parallel universes, we obtain a single de-Sitter space bounded between two Randall-Sundrum branes.

A multiverse model is obtained by taking many slices of a single de-Sitter space as done in \cite{Yadav:2023qfg} for the AdS bulk spacetime. In the current work, the Multiverse consists of many universes (each universe with de-Sitter geometry) joined at the common defect, and all of these universes can communicate with each other via the transparent boundary conditions at the defect. These universes are embedded in higher dimensional de-Sitter space. Interestingly, we can obtain connected multiverses by gluing many copies of a single multiverse model. However, in the connected multiverses, if all the universes of a multiverse are glued with their consecutive Multiverse, then we obtain a single de-Sitter space bounded by two end-of-the-world branes.

The rest of the paper is organized as follows. We start with the construction of wedge holography for the de-Sitter bulk in section \ref{WHMD} via three subsections \ref{sec-2.1}, \ref{WHVD} and \ref{PR-WHDS}. In subsection \ref{sec-2.1}, we describe the mathematical description of de-Sitter wedge holography; in subsection \ref{WHVD}, we prove that we have consistent wedge holographic realization for the de-Sitter bulk, and in subsection \ref{PR-WHDS}, we describe de-Sitter wedge holography pictorially. We construct the wedge holography in an extended static patch in section \ref{WH-ESP-sec} and use this model to discuss whether parallel universes exist or not in section \ref{PU-ESP}. In section \ref{PM-dSWH}, we discuss the existence of multiverse (subsection \ref{Multiverse-sec}) and communicating multiverses (subsection \ref{PM-subsec}). We make comments on de-Sitter holography in section \ref{dS-CFT correspondance} and then discuss our results in section \ref{conclusion}.

\section{Wedge holography for the de-Sitter bulk: general description}
\label{WHMD}
In this section, we construct wedge holography for the general de-Sitter bulk and prove that the metric on the end-of-the-world branes\footnote{At many place we will use ``EOW branes'' instead of ``end-of-the-world branes''.} can be static patch and global coordinate. We consider dimensionality of de-Sitter bulk as $d$-dimensions whereas dimensionality of end-of-the-world branes [located at $\omega=\omega_1$ and $\omega=\omega_2$] as $(d-1)$-dimensions. The discussion in this section will be general, similar to wedge holography construction for the AdS bulk.

\subsection{The setup}
\label{sec-2.1}
The bulk gravitational action with positive cosmological constant in $d$-dimensions including the gravitational action on the branes is given as \cite{Geng:2021wcq}
\begin{eqnarray}
\label{Bulk-action}
S=\frac{1}{16 \pi G_N^{(d)}}\int d^d x \sqrt{-g} { \left({\cal R}- 2 \Lambda \right)}-\frac{1}{8 \pi G_N^{(d)}}\int d^{d-1} x \sqrt{-\hat{g}_{\beta}}\left({\cal K}_{\beta}-T_{\beta}\right),
\end{eqnarray}
where $\beta=1,2$ and $\Lambda=\frac{(d-1)(d-2)}{2}$. $G_N^{(d)}$, ${\cal R}$ and $\Lambda$ are the Newton constant, Ricci scalar and cosmological constant in $d$-dimensions; ${\cal K}_{\beta}$ and $T_{\beta}$ are the trace of extrinsic curvature and tensions of end-of-the-world branes. Einstein equation associated with the bulk metric is given by
\begin{eqnarray}
\label{Bulk-Einstein-Eqn}
 R_{\mu\nu}-\frac{R}{2}g_{\mu\nu}{\bf +}\Lambda g_{\mu\nu}=0.
\end{eqnarray}
The above equation has the following solution \cite{Alishahiha:2004md}
\begin{eqnarray}
\label{bulk-metric}
& & 
ds^2= d \omega^2+e^{2 A(\omega)}\hat{g}_{ij} dy^i dy^j,
\end{eqnarray}
where $\hat{g}_{ij}$ corresponds to the metric of dS$_{d-1}$ space. Depending upon the warp factor, the bulk metric (\ref{bulk-metric}) can be either de-Sitter or anti de-Sitter\footnote{Bulk action: \begin{eqnarray}
\label{Bulk-action-AdS}
S=-\frac{1}{16 \pi G_N^{(d)}}\int d^d x \sqrt{-g} { \left({\cal R}- 2 \Lambda \right)}-\frac{1}{8 \pi G_N^{(d)}}\int d^{d-1} x \sqrt{-\hat{g}_{\beta}}\left({\cal K}_{\beta}-T_{\beta}\right),
\end{eqnarray}
with $\Lambda=-\frac{(d-1)(d-2)}{2}$.} as follows \cite{Alishahiha:2004md}
\begin{eqnarray}
\label{warp-factors}
& & {\bf dS_d}: \ \  \ \ \  \ \ \  \ e^{2 A(\omega)} = \sin^2\left(\omega/R\right), \nonumber\\
& & {\bf AdS_d}: \ \  \ \ \  \  \  e^{2 A(\omega)} = \sinh^2\left(\omega/R\right),
\end{eqnarray}
where $R$ is the length scale associated with corresponding background. Since we are interested in constructing wedge holography for the de-Sitter bulk, we will consider the de-Sitter case here. Let us write the explicit metric for the de-Sitter bulk as 
\begin{eqnarray}
\label{bulk-metric-dS}
& & 
 ds_{\rm general \ bulk}^2= d \omega^2+\sin^2\left(\omega/R\right)\hat{g}_{ij}^\beta dy^i dy^j,\nonumber\\
& & ds_{\rm static \ patch}^2 =d \omega^2+\sin^2\left(\omega/R\right)\left[-(1-r^2)dt^2+\frac{dr^2}{1-r^2}+r^2 d \Omega_{d-3}^2 \right], \nonumber\\
& & ds_{\rm global \ de-Sitter}^2 =d \omega^2+ \sin^2\left(\omega/R\right) \left[-d \tau^2+\cosh^2\left(\tau\right) d \Omega_{d-2}^2 \right],
\end{eqnarray}
where $t \in (-\infty,\infty)$,  $r \in (0,1)$\footnote{We set $l=1$ and the horizon is located at $r=1$.}, $\tau \in (-\infty,\infty)$, and \\$d \Omega_{d-2}^2=d\theta_1^2+ \sin^2\theta_1 d\theta_2^2+......+\sin^2\theta_1 \sin^2\theta_2......\sin^2\theta_{d-3} d\theta_{d-2}^2$. \par
The Neumann boundary condition (NBC) satisfied by the bulk metric (\ref{bulk-metric}) at the location of end-of-the-world branes is given as
\begin{eqnarray}
\label{NBC}
{\cal K}_{ij}^{\beta}-({\cal K}^{\beta}-T^{\beta})\hat{g}_{ij}^{\beta}=0,
\end{eqnarray}
where ${\cal K}_{ij}^{\beta}=\frac{1}{2} \partial_\omega {\hat{g}_{ij}^{\beta}}|_{\omega= \omega_{1,2}}$ and ${\cal K}^\beta=\hat{g}^{ij,\beta}{\cal K}_{ij,\beta}$.\par

It has been discussed in \cite{Alishahiha:2004md} that the de-Sitter bulk (\ref{bulk-metric-dS}) is dual to two CFTs on $dS_{d-1}$ which are coupled to each other and $(d-1)$-dimensional de-Sitter gravity at $\omega=\frac{\pi R}{2}$. The warp factor is maximum at $\omega=\frac{\pi R}{2}$ and it is zero at the horizon, $\omega=0,\pi R$. For the discussion in this section, we will consider localization of $d$-dimensional de-Sitter gravity on $(d-1)$-dimensional EOW brane \cite{Karch-dS}.\par

Having discussed the general structure of $d$-dimensional de-Sitter space, let us see how we can construct wedge holography in de-Sitter space. Let us consider two EOW branes located at $\omega=\omega_1$ and $\omega=\omega_2$ which are basically constant $\omega$ slicing of the bulk metric (\ref{bulk-metric-dS}). It is to be noted that the story here is different from the AdS bulk because wedge holography construction of de-Sitter bulk is a little bit difficult. We will discuss this pictorially in subsection \ref{PR-WHDS} using the idea of \cite{Kawamoto:2023wzj}. For now, we are considering EOW branes at some arbitrary locations $\omega_{1,2}$. The wedge holography is obtained by joining two $(d-1)$-dimensional EOW branes (say $Y_1$ and $Y_2$) with de-Sitter metric at the $(d-2)$-dimensional defect in $d$-dimensional de-Sitter bulk (see Fig. \ref{WH-single-dS}). \par

Similar to AdS and flat bulk spacetime, we expect that de-Sitter wedge holography has the following three descriptions:
\begin{itemize}
\item {\bf Boundary description}: $(d-2)$-dimensional defect CFT living at the corner of the wedge formed by two EOW branes.

\item {\bf Intermediate description}: the gravitating branes ($Y_1$ and $Y_2$) with metric $dS_{d-1}$ are interacting with each other via the transparent boundary conditions at the defect.

\item {\bf Bulk description}: CFT living on the $(d-2)$-dimensional defect is dual to $d$-dimensional de-Sitter gravity living in the wedge formed by gravitating EOW branes with metric dS$_{d-1}$.
\end{itemize}
When we take de-Sitter slicing of (\ref{bulk-metric}) then we expect that defect CFT should be non-unitary because of dS/CFT correspondence \cite{Strominger:2001pn,dS-CFT-1}. Let us make statement about the wedge holography  \cite{WH-1,WH-2} in the context of de-Sitter bulk as follows:\\

\fbox{\begin{minipage}{38em}{\it 
Classical gravity in $d$-dimensional de-Sitter space \\ $\equiv$ (Quantum) gravity on two $(d-1)$-dimensional EOW branes with metric dS$_{d-1}$\\ $\equiv$ CFT living at the $(d-2)$-dimensional defect.}
\end{minipage}}\\ \\
The first and the second lines are related via braneworld holography in de-Sitter space \cite{Karch-dS} whereas the second and third lines exist because of dS/CFT correspondence \cite{Strominger:2001pn,dS-CFT-1}. Hence,  {\it classical gravity in $d$-dimensional de-Sitter space is dual to $(d-2)$-dimensional defect CFT living at the corner of the wedge, which is the signature of dS$_d$/CFT$_{d-2}$ correspondence.} 

To get a consistent wedge holographic realization of the de-Sitter space, we need to check the following conditions:
\begin{enumerate}
\item The bulk metric (\ref{bulk-metric-dS}) should satisfy Einstein's equation with a positive cosmological constant (\ref{Bulk-Einstein-Eqn}).

\item The bulk metric (\ref{bulk-metric-dS}) must satisfy Neumann boundary conditions (\ref{NBC}) on the branes, i.e. at $\omega=\omega_{1,2}$ where $\omega_{1,2}$ are some constant values of $\omega$.

\item The induced metric $\hat{g}_{ij}^\beta$ on the branes should be the solution of Einstein's equation with a positive cosmological constant in $(d-1)$-dimensions:
\begin{eqnarray}
\label{Brane-Einstein-equation-dS}
R_{ij}^\beta-\frac{1}{2}\hat{g}_{ij}^\beta R[\hat{g}_{ij}^\beta]+\frac{(d-2)(d-3)}{2} \hat{g}_{ij}^\beta=0.
\end{eqnarray}
\end{enumerate}
One can obtain the gravitational action on the branes from the substitution of (\ref{bulk-metric-dS}) into the bulk action (\ref{Bulk-action}), using the Neumann boundary condition (\ref{NBC}) and integration over $\omega$ and the general form is given as
\begin{eqnarray}
\label{brane-action-dS}
S_{\rm brane}= \lambda_\beta^{\rm dS}\int d^{d-1}x \left(R[\hat{g}_{ij}^\beta]-2 \Lambda_{\rm brane}^{\rm dS}\right),
\end{eqnarray}
where $\Lambda_{\rm brane}^{\rm dS}=\frac{(d-2)(d-3)}{2}$, $\lambda_\beta^{\rm dS}=\frac{1}{16 \pi G_N^{(d-1),\beta}}$. Here, we show that condition {\bf 2} is satisfied, provided EOW branes have some tensions. Let us see this in more detail. For the bulk metric (\ref{bulk-metric-dS}), extrinsic curvature and the trace of the same on EOW branes are obtained as
\begin{eqnarray}
\label{Extrinsic-curvature-dS-bulk}
& & {\cal K}_{ij}^{\beta}=\frac{1}{2} \partial_\omega {\hat{g}_{ij}}|_{\omega= \omega_{1,2}}=\frac{1}{R} \cot\left(\frac{\omega_{1,2}}{R} \right)\hat{g}_{ij}|_{\omega= \omega_{1,2}},\nonumber\\
& &  {\cal K}^\beta=\hat{g}^{ij,\beta} {\cal K}_{ij}^{\beta}=\frac{d-1}{R} \cot\left(\frac{\omega_{1,2}}{R} \right).
\end{eqnarray}
Hence, using the Neumann boundary condition (\ref{NBC}) and the information given in (\ref{Extrinsic-curvature-dS-bulk}), we see that NBC for the bulk metric (\ref{bulk-metric-dS}) is satisfied when tensions of the branes are given by 
\begin{eqnarray}
\label{T-dS}
T^\beta=\frac{d-2}{R} \cot\left(\frac{\omega_{1,2}}{R} \right).
\end{eqnarray} 
Hence at $\omega_{1,2}=\frac{\pi R}{2}$, we get tensionless branes as discussed in \cite{Geng:2021wcq}.

\subsection{Wedge holography in various dimensions}
\label{WHVD}
In this subsection, we construct wedge holography for various dimensional de-Sitter bulk. For simplicity, we work with $R=1$ unit. 
\subsubsection{$d=3$}
For three-dimensional de-Sitter space, the bulk metric is given as
\begin{eqnarray}
\label{bulk-metric-dS3}
& & 
ds^2=d \omega^2+\sin^2\left(\omega\right)\hat{g}_{ij}^\beta dy^i dy^j,\nonumber\\
& & ds_{\rm static \ patch}^2 =d \omega^2+\sin^2\left(\omega\right)\left[-(1-r^2)dt^2+\frac{dr^2}{1-r^2} \right], \nonumber\\
& & ds_{\rm global \ de-Sitter}^2 = d \omega^2+ \sin^2\left(\omega\right) \left[-d \tau^2+\cosh^2\left(\tau\right) d \Omega_{1}^2 \right],
\end{eqnarray}
where $d \Omega_{1}^2=d\theta_1^2$. For this case, the Ricci scalar is $R=6$, and the Ricci tensors for the static patch and global de-Sitter are summarized below.
\begin{itemize}
\item {\bf Ricci tensors for static patch}: 
\begin{eqnarray}
\label{RTs-dS3-SP}
& & R_{11}=2 \left(r^2-1\right) \sin ^2(w); \ \ R_{22}=-\frac{2 \sin ^2(\omega)}{r^2-1}; \ \ R_{33}=2.
\end{eqnarray}

\item {\bf Ricci tensors for global coordinate}: 
\begin{eqnarray}
\label{RTs-dS3-GC}
& & R_{11}=-2 \sin ^2(\omega); \ \ R_{22}=2 \cosh^2(\tau) \sin ^2(\omega); \ \ R_{33}=2.
\end{eqnarray}
\end{itemize}
The cosmological constant for three-dimensional de-Sitter space is $\Lambda=1$, and hence bulk EOM is given as
\begin{eqnarray}
\label{bulk-EOM-dS3}
R_{\mu \nu}-\frac{1}{2}g_{\mu \nu}R+g_{\mu \nu}=0,
\end{eqnarray}
where $\mu,\nu=1,2,3$.
One can check that bulk EOM\footnote{EOM is the short form of equation of motion.} (\ref{bulk-EOM-dS3}) is satisfied for the Ricci tensors given in equations (\ref{RTs-dS3-SP}) and (\ref{RTs-dS3-GC}) and the Ricci scalar obtained above. Hence, condition {\bf 1} is satisfied for bulk dS$_3$. Therefore, we get dS$_2$ slicing of the bulk metric (\ref{bulk-metric-dS3}) in the static patch and global coordinate both. Further, the tensions of the end-of-the-world branes and trace of the extrinsic curvature for dS$_3$ bulk are given as
\begin{eqnarray}
\label{T-K-dS3}
T^\beta= \cot\left(\omega_{1,2}\right); \ \ \ {\cal K}^\beta=2 \cot\left(\omega_{1,2}\right).
\end{eqnarray}

\subsubsection{$d=4$}
The metric for the four-dimensional de-Sitter space is given by
\begin{eqnarray}
\label{bulk-metric-dS-d=4}
& & 
ds^2= d \omega^2+\sin^2\left(\omega\right)\hat{g}_{ij}^\beta dy^i dy^j,\nonumber\\
& &ds_{\rm static \ patch}^2 =d \omega^2+\sin^2\left(\omega\right)\left[-(1-r^2)dt^2+\frac{dr^2}{1-r^2}+r^2 d \Omega_{1}^2 \right], \nonumber\\
& &ds_{\rm global \ de-Sitter}^2 =d \omega^2+ \sin^2\left(\omega\right) \left[-d \tau^2+\cosh^2\left(\tau\right) d \Omega_{2}^2 \right],
\end{eqnarray}
where $d \Omega_{2}^2=d\theta_1^2+\sin^2\theta_1 d\theta_2^2$ and the Ricci tensors for the static patch and global de-Sitter are given as below.
\begin{itemize}
\item {\bf Static patch}: 
\begin{eqnarray}
\label{RTs-SP-dS4}
& &  \hskip -0.2in R_{11}=3 \left(r^2-1\right) \sin ^2(\omega); \ \ R_{22}=-\frac{3 \sin ^2(\omega)}{r^2-1}; \ \ R_{33}=3 r^2 \sin ^2(\omega); \ \ R_{44}=3.
\end{eqnarray}

\item {\bf Global coordinate}: 
\begin{eqnarray}
\label{RTs-GC-dS4}
& & \hskip -0.5in R_{11}=-3 \sin ^2(\omega); \ \ R_{22}=3 \cosh^2(\tau) \sin ^2(\omega); \ \ R_{33}=3 \sin^2\left(\theta _1\right) \cosh^2(\tau) \sin ^2(\omega); \ \ R_{44}=3. \nonumber\\
\end{eqnarray}
\end{itemize}
The Ricci scalar for dS$_4$ metric (\ref{bulk-metric-dS-d=4}) is $R=12$ and the cosmological constant is, $\Lambda=3$ and hence the EOM for the bulk metric associated with dS$_4$ space is
\begin{eqnarray}
\label{EOM-bulk-dS4}
R_{\mu \nu}-\frac{1}{2}g_{\mu \nu}R+3 g_{\mu \nu}=0,
\end{eqnarray}
where $\mu,\nu=1,2,3,4$. We see that (\ref{EOM-bulk-dS4}) is satisfied for the Ricci tensors given in (\ref{RTs-SP-dS4}) and (\ref{RTs-GC-dS4}) and the Ricci scalar as mentioned above which proves condition {\bf 1} of wedge holography for the dS$_4$ bulk. For the dS$_4$ bulk, tensions of the end-of-the-world branes and trace of the extrinsic curvature are
\begin{eqnarray}
\label{T-K-dS4}
T^\beta= 2 \cot\left(\omega_{1,2}\right); \ \ \ {\cal K}^\beta=3 \cot\left(\omega_{1,2}\right).
\end{eqnarray}

\subsubsection{$d=5$}
The five-dimensional bulk de-Sitter metric is written as
\begin{eqnarray}
\label{bulk-metric-dS-d=5}
& & 
ds^2= d \omega^2 +\sin^2\left(\omega\right)\hat{g}_{ij}^\beta dy^i dy^j,\nonumber\\
& & ds_{\rm static \ patch}^2 =d \omega^2+\sin^2\left(\omega\right)\left[-(1-r^2)dt^2+\frac{dr^2}{1-r^2}+r^2 d \Omega_{2}^2 \right] , \nonumber\\
& & ds_{\rm global \ de-Sitter}^2 =d \omega^2+ \sin^2\left(\omega\right) \left[-d \tau^2+\cosh^2\left(\tau\right) d \Omega_{3}^2 \right],
\end{eqnarray}
where $d \Omega_{3}^2=d\theta_1^2+\sin^2\theta_1 d\theta_2^2+\sin^2\theta_1 \sin^2\theta_2 d\theta_3^2$. The Ricci tensors are given below.
\begin{itemize}
\item {\bf Static patch}: 
\begin{eqnarray}
& & R_{11}=4 \left(r^2-1\right) \sin ^2(\omega); \ \ R_{22}=-\frac{4 \sin ^2(\omega)}{r^2-1}; \ \ R_{33}=4 r^2 \sin ^2(\omega); \nonumber\\
& &  R_{44}=4 r^2 \sin ^2\left(\theta _1\right) \sin ^2(\omega); \ \ R_{55}=4.
\end{eqnarray}

\item {\bf Global coordinate}: 
\begin{eqnarray}
& &  R_{11}=-4 \sin ^2(\omega); \ \ R_{22}=4 \cosh^2(\tau) \sin ^2(\omega); \ \ R_{33}=4 \sin ^2\left(\theta _1\right) \cosh^2(\tau) \sin ^2(\omega); \nonumber\\
& &  R_{44}=4 \sin ^2\left(\theta _1\right) \sin ^2\left(\theta _2\right) \cosh^2(\tau) \sin ^2(\omega); \ \ R_{55}=4. \nonumber\\
\end{eqnarray}
\end{itemize}
Ricci scalar for dS$_5$ bulk is  $R=20$, and the cosmological constant for the same is $\Lambda=6$. Hence bulk EOM is given as
\begin{eqnarray}
\label{EOM-dS5}
R_{\mu \nu}-\frac{1}{2}g_{\mu \nu}R+6 g_{\mu \nu}=0,
\end{eqnarray}
where $\mu,\nu=1,2,3,4,5$. Again we can check that bulk EOM (\ref{EOM-dS5}) is satisfied for the Ricci tensors and Ricci scalar mentioned above for the dS$_5$ bulk for the static patch as well as global coordinate metric on EOW branes and hence condition {\bf 1} is satisfied. 

\subsubsection{$d=6$}
For the six-dimensional de-Sitter space, the bulk metric is given as
\begin{eqnarray}
\label{bulk-metric-dS-d=6}
& & 
ds^2= d \omega^2+\sin^2\left(\omega\right)\hat{g}_{ij}^\beta dy^i dy^j,\nonumber\\
& & ds_{\rm static \ patch}^2  =d \omega^2+\sin^2\left(\omega\right)\left[-(1-r^2)dt^2+\frac{dr^2}{1-r^2}+r^2 d \Omega_{3}^2 \right], \nonumber\\
& & ds_{\rm global \ de-Sitter}^2 = d \omega^2+\sin^2\left(\omega\right) \left[-d \tau^2+\cosh^2\left(\tau\right) d \Omega_{4}^2 \right],
\end{eqnarray}
where $d \Omega_{4}^2=d\theta_1^2+\sin^2\theta_1 d\theta_2^2+\sin^2\theta_1 \sin^2\theta_2 d\theta_3^2+\sin^2\theta_1 \sin^2\theta_2 \sin^2\theta_3 d\theta_4^2$ and the Ricci tensors are written below.
\begin{itemize}
\item {\bf Static patch}: 
\begin{eqnarray}
& & R_{11}=5 \left(r^2-1\right) \sin ^2(\omega); \ \ R_{22}=-\frac{5 \sin ^2(\omega)}{r^2-1}; \ \ R_{33}=5 r^2 \sin ^2(\omega); \nonumber\\
& &  R_{44}=5 r^2 \sin ^2\left(\theta _1\right) \sin ^2(\omega); \ \ R_{55}=5 r^2 \sin ^2\left(\theta _1\right) \sin ^2\left(\theta _2\right) \sin ^2(\omega); \ \ R_{66}=5.
\end{eqnarray}

\item {\bf Global coordinate}: 
\begin{eqnarray}
& &  R_{11}=-5 \sin ^2(\omega); \ \ R_{22}=5 \cosh^2(\tau) \sin ^2(\omega); \ \ R_{33}=5 \sin ^2\left(\theta _1\right) \cosh^2(\tau) \sin ^2(\omega); \nonumber\\
& &  R_{44}=5 \sin ^2\left(\theta _1\right) \sin ^2\left(\theta _2\right) \cosh^2(\tau) \sin ^2(\omega); \nonumber\\
& &  R_{55}=5 \sin ^2\left(\theta _1\right) \sin ^2\left(\theta _2\right) \sin ^2\left(\theta _3\right) \cosh^2(\tau) \sin
   ^2(\omega); \ \ R_{66}=5.
\end{eqnarray}
\end{itemize}
Ricci scalar for dS$_6$ is $R=30$, and the cosmological constant for dS$_6$ is $\Lambda=10$, which leads to the following EOM for the bulk metric (\ref{bulk-metric-dS-d=6})
\begin{eqnarray}
\label{EOM-dS6}
R_{\mu \nu}-\frac{1}{2}g_{\mu \nu}R+10 g_{\mu \nu}=0,
\end{eqnarray}
where $\mu,\nu=1,2,3,4,5,6$. Similar to other cases discussed earlier, it is easy to check that the bulk metric (\ref{bulk-metric-dS-d=6}) satisfies (\ref{EOM-dS6}).

To summarize, we proved that one could construct wedge holography for the de-Sitter bulk, and we have done this in $d=3,4,5,6$, and in general, it is valid for arbitrary $d$-dimensional de-Sitter space. Now, we are done with the mathematical description of wedge holography in de-Sitter space, and hence in subsection \ref{PR-WHDS}, we will discuss the pictorial realization of the same and other related issues.

\subsection{Pictorial representation of wedge holography for the de-Sitter bulk}
\label{PR-WHDS}
\begin{figure}[h]
  \centering
  \includegraphics[width=.4\linewidth]{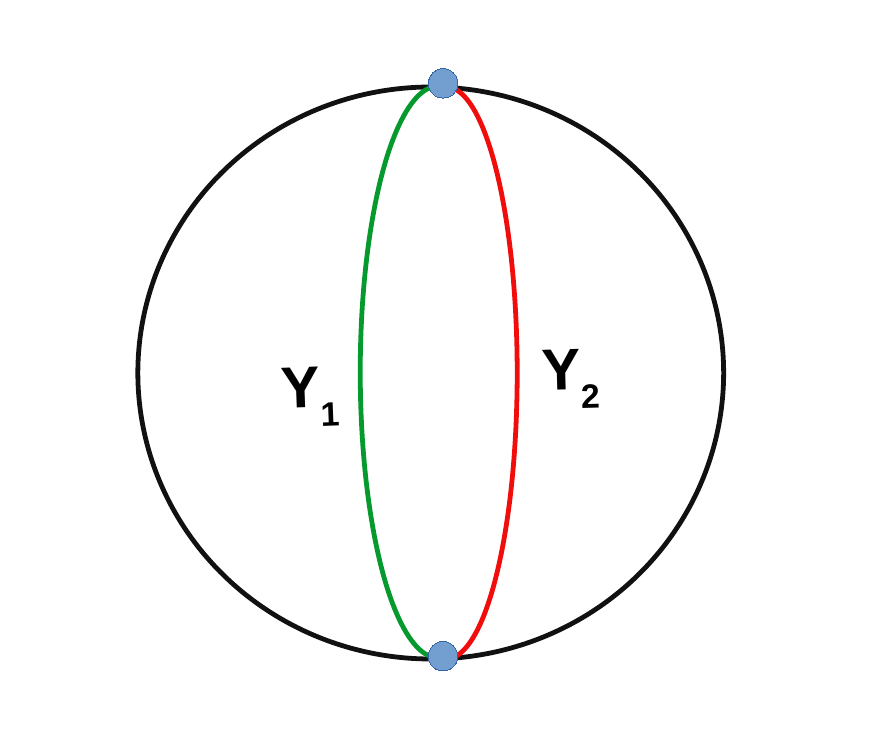}
\caption{Pictorial representation of wedge holography in a single de-Sitter space. Green and red curves denote EOW branes at $\omega=\omega_1$ ($Y_1$) and $\omega=\omega_2$ ($Y_2$). Blue dots are $(d-2)$-dimensional defects.}
\label{WH-single-dS}
\end{figure}

We can construct the wedge holography from a single de-Sitter bulk by considering end-of-the-world branes at constant $\omega$ slices as done in \cite{GB-3,Yadav:2023sdg}. For example, we have to consider $Y_1$ at $\omega=\omega_1$ and $Y_2$ at $\omega=\omega_2$ as shown in Fig. \ref{WH-single-dS} where we have shown the constant $t$ slice of de-Sitter space.
\par
\begin{figure}[h]
  \centering
  \includegraphics[width=.8\linewidth]{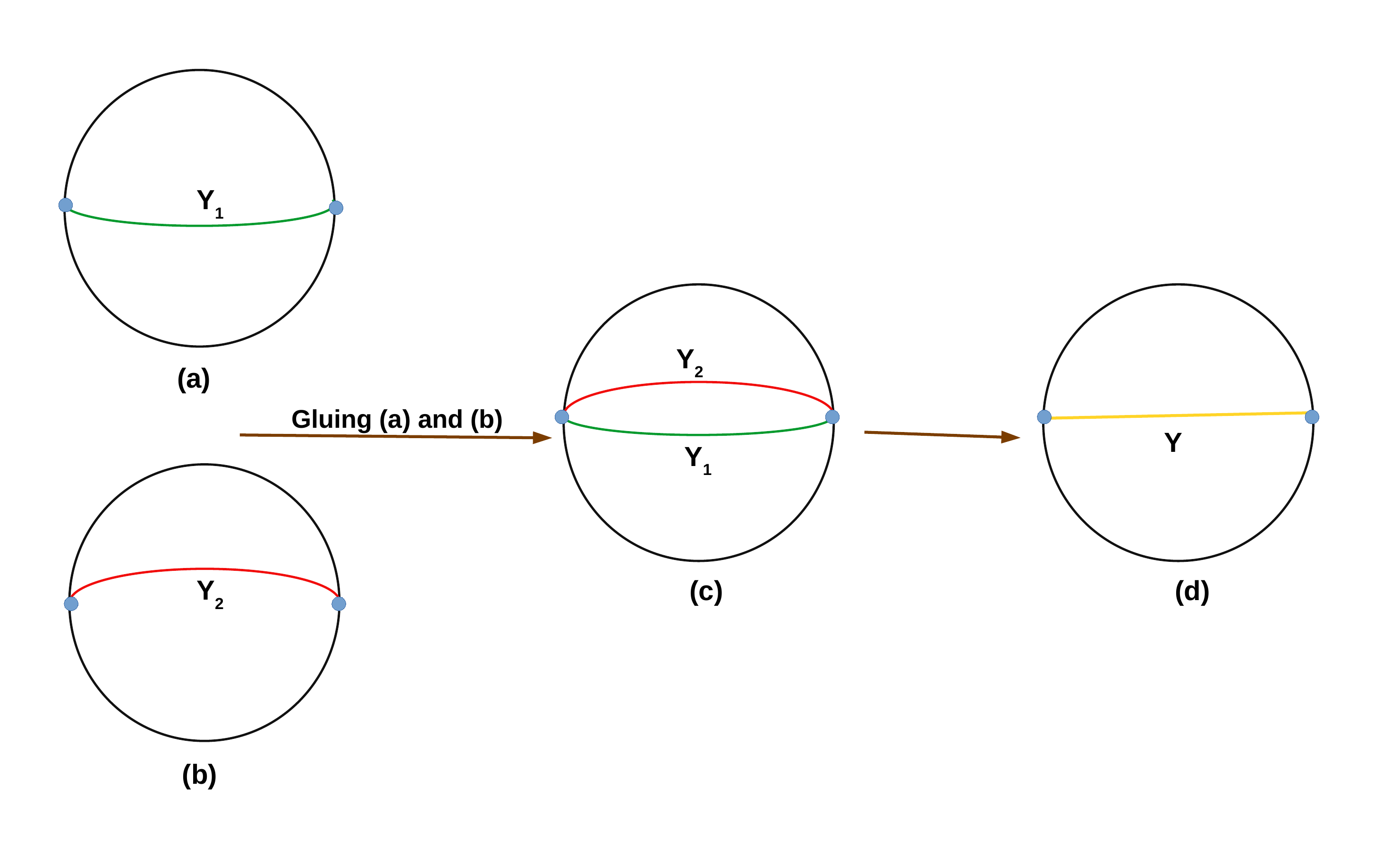}
\caption{Pictorial representation of wedge holography in de-Sitter space. Green and red curves denote EOW branes at $\omega=\omega_1$ ($Y_1$) and $\omega=\omega_2$ ($Y_2$). We can glue (a) and (b) \cite{Kawamoto:2023wzj} which results in (c) and (d).}
\label{WH}
\end{figure}

{\bf Wedge holography by joining two copies of double holography in de-Sitter space}: Here, we discuss the idea that will play an important role in constructing extended static patch wedge holography. The wedge holography in de-Sitter space can also be described pictorially, as shown in Fig. \ref{WH}. 
We take the spatial slice of de-Sitter space and take a constant $\omega$ slice at $\omega=\omega_1$; this is analogous to chopping off the geometry beyond $\omega=\omega_1$ (description (a) in Fig. \ref{WH}). We call this slice as $Y_1$. Similarly, we take one more copy of the setup above and chop off the geometry at $\omega=\omega_2$; we name this constant $\omega$ slice as $Y_2$ (description (b) in Fig. \ref{WH}). Now we join (a) and (b) using \cite{Kawamoto:2023wzj}. {Conditions for gluing $Y_1$ (with metric $\hat{g}_{ij}^{(1)}$, tension $T^{(1)}$) and $Y_2$ (with metric $\hat{g}_{ij}^{(2)}$, tension $T^{(2)}$) are Israel junction conditions\footnote{This condition has to be satisfied whenever we join two spacetimes along the two branes.}
\begin{eqnarray}
& & \hat{g}_{ij}^{(1)}=\hat{g}_{ij}^{(2)}; \ \ \ \ [K_{ij}]-[K]\hat{g}_{ij}=- (T^{(1)}+T^{(2)}) \hat{g}_{ij}, \nonumber
\end{eqnarray}
where $[K_{ij}]=K_{ij}^{(1)}+K_{ij}^{(2)}$. 
 In the end, $Y_1$ and $Y_2$ will merge and result in a new brane, say ``$Y$'', and we see that we joined two de-Sitter spaces and obtained a new de-Sitter space.
\par
Therefore, if we compare the above construction of the wedge holography in de-Sitter space with the usual construction in AdS bulk, then we see that we can see a wedge holography-like structure as long as $Y_1$ and $Y_2$ are separated. Whenever two EOW branes $Y_1$ and $Y_2$ merge into a single brane $Y$, we obtain a single de-Sitter space instead of wedge holography.

In the next section, we will construct wedge holography for the de-Sitter bulk for the extended static patch, including two Randall-Sundrum branes \cite{Randall:1999vf,Randall:1999ee}. In that case, wedge holography is constructed by two Randall-Sundrum branes $Q_1$ and $Q_2$, similar to the wedge holography in AdS bulk.

\section{Wedge holography in extended static patch}
\label{WH-ESP-sec}
Authors in \cite{Geng:2021wcq} constructed double holography in an extended static patch in the context of DS/dS correspondence. The bulk metric of ($d+1$)-dimensional de-Sitter space is
\begin{eqnarray}
\label{metric-ESP}
ds_{d+1}^2=d \omega^2+\sin^2(\omega)ds_d^2=d \omega^2+\sin^2(\omega)\left(-\cos^2\beta dt^2+d\beta^2+\sin^2\beta d\Omega_{d-2}^2\right),
\end{eqnarray}
where $\beta \in (0,\pi)$ for $d\geq 3$ and $d\Omega_{d-2}^2=d\chi^2 +\sin^2 \chi d\Omega_{d-3}^2$ such that $0 \leq \chi \leq \pi$ for $d \geq 4$.
The idea to construct double holography is that we need to consider two d-dimensional de-Sitter branes, dS$_d^1$ and dS$_d^2$ and treat dS$_d^1$ brane as Randall-Sundrum brane $Q$ (described by $\chi=\pi/2$ slice) and take the $\mathbb{Z}_2$ quotient of dS$_d^2$, i.e., dS$_d^2/\mathbb{Z}_2$ which removes half of de-Sitter space of dS$_d^2$ and its spatial slice looks like hemisphere. In \cite{Geng:2021wcq}, dS$_d^2/\mathbb{Z}_2$ is described by $\omega =\pi/2$ slice and is treated as a bath analogous to double holography in AdS backgrounds. In de-Sitter double holography, $d$-dimensional Randall-Sundrum brane $Q$ interact with dS$_d^2/\mathbb{Z}_2$ via $(d-1)$-dimensional defect where authors considered transparent boundary conditions.

Since $\omega=\pi/2$ slice is the UV part of the geometry \cite{Alishahiha:2004md} therefore, we can treat dS$_d^2/\mathbb{Z}_2$ as the UV brane. By following \cite{Kawamoto:2023wzj}, we can glue two copies of the double holography described above, and this is described pictorially in Fig. \ref{WH-ESP}.
\begin{figure}[h]
  \centering
  \includegraphics[width=.9\linewidth]{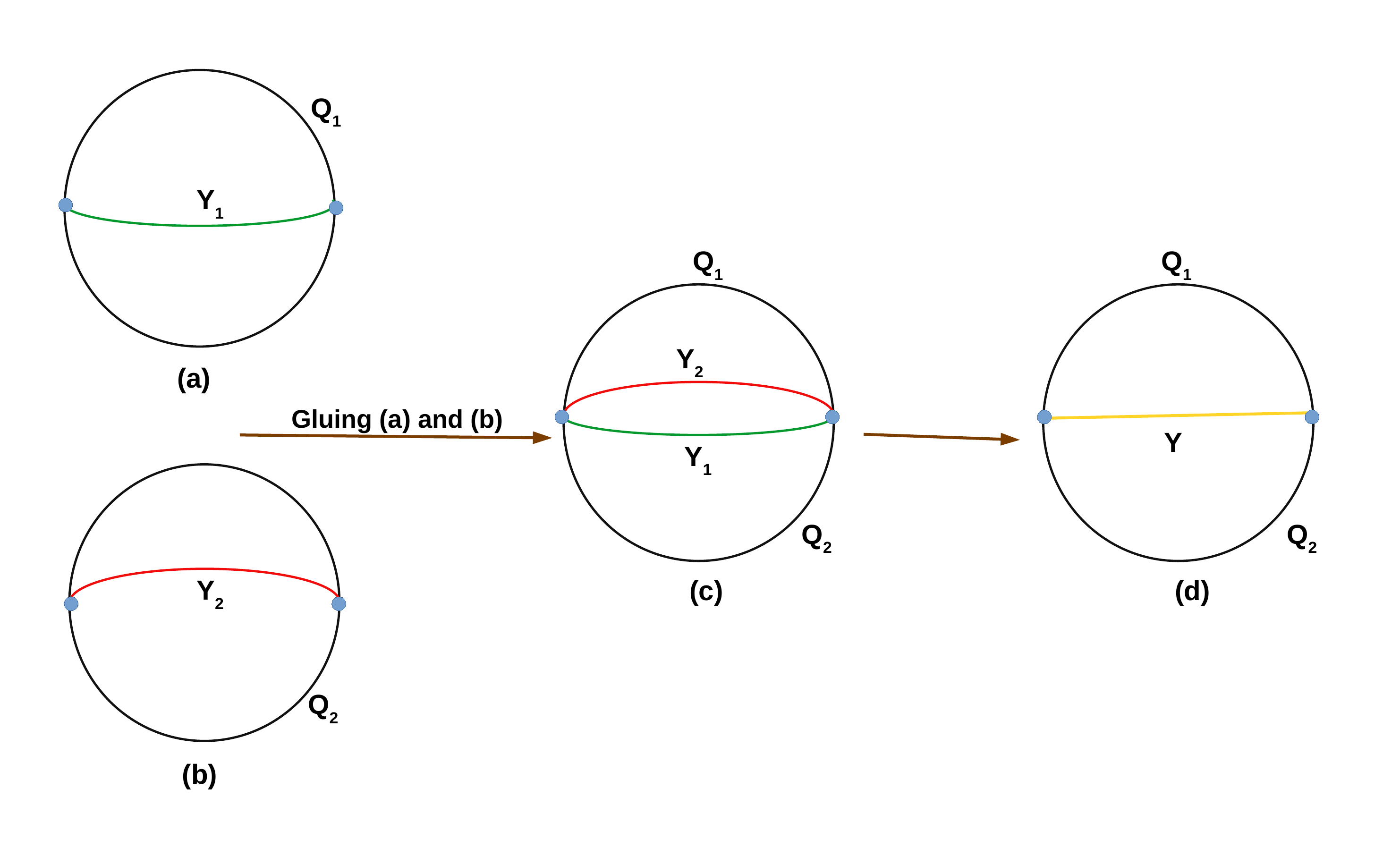}
\caption{Realization of wedge holography in DS/dS correspondence. Green and red curves denote UV branes $Y_1$ and  $Y_2$ at $\omega_{1,2}=\pi/2$. (a) and (b) are joined along the branes $Y_1$ and  $Y_2$ which results in the final picture (d).}
\label{WH-ESP}
\end{figure}
\par
Mathematical descriptions of gluing two copies of double holography is discussed below.
\begin{itemize}
\item Consider the bulk metric as
\begin{eqnarray}
\label{metric-ESP-i}
& & ds_{d+1}^2=d \omega_1^2+\sin^2(\omega_1)dS_d^2\nonumber\\
& & \hskip 0.4in =d \omega_1^2+\sin^2(\omega_1)\left(-\cos^2\beta dt^2+d\beta^2+\sin^2\beta \left(d\chi_1^2 +\sin^2{\chi_1} d\Omega_{d-3}^2\right)\right).
\end{eqnarray}
In geometry (\ref{metric-ESP-i}), we consider two slicing, $\omega_1=\pi/2$ [dS$_d^2/\mathbb{Z}_2$, we denote this as $Y_1$] and $\chi=\pi/2$ [Randall-Sundrum brane, $Q_1$ which is dS$_d^1$]. $Q_1$ interact with $Y_1$ via the transparent boundary conditions at the $(d-1)$-dimensional defect.

\item Let us take the bulk metric as
\begin{eqnarray}
\label{metric-ESP-ii}
& & ds_{d+1}^2=d \omega_2^2+\sin^2(\omega_2)dS_d^2
\nonumber\\
& & \hskip 0.4in=d \omega_2^2+\sin^2(\omega_2)\left(-\cos^2\beta dt^2+d\beta^2+\sin^2\beta \left(d\chi_2^2 +\sin^2{\chi_2} d\Omega_{d-3}^2\right)\right).
\end{eqnarray}
In the bulk (\ref{metric-ESP-ii}), $\omega_2=\pi/2$ slice is dS$_d^4/\mathbb{Z}_2$ ($Y_2$) and $\chi=\pi/2$ slice is dS$_d^3$ (Randall-Sundrum brane, $Q_2$). The interaction between $Q_2$ and $Y_2$ is happening due to the transparent boundary conditions at the $(d-1)$-dimensional defect. 

\end{itemize}
In double holography, the space available is between the Randall-Sundrum brane $Q$ and UV brane $Y$. {\it Wedge holography in the extended static patch is constructed by two Randall-Sundrum branes $Q_1$ and $Q_2$ (description (d) in Fig. \ref{WH-ESP}). We have ($d+1$)-dimensional de-Sitter gravity in the bulk, which is localized on the $d$-dimensional Randall-Sundrum branes $Q_1$ and $Q_2$.
Joining of two Randall-Sundrum branes $Q_1$ and $Q_2$ at the ($d-1$)-dimensional defect (blue dots in description (d) of Fig. \ref{WH-ESP}) form a wedge.} In the language of \cite{Kawamoto:2023wzj}, we have a de-Sitter space bounded between $Q_1$ and $Q_2$.

\section{Communicating universes in extended static patch wedge holography?}
\label{PU-ESP}
The setup described in section \ref{WH-ESP-sec} can describe the communicating universes similar to \cite{Aguilar-Gutierrez:2023zoi}\footnote{We can also call this kind of setup as ``parallel universe'' because all the universes are parallel to each other \cite{Kaku}. All de-sitter branes are not joined at the common defect. See \cite{Tegmark:2003sr} for the classification of parallel universes.}. We consider ``$n$'' copies of the wedge holography discussed in section \ref{WH-ESP-sec} to describe communicating universes. Communicating universes are obtained by joining $Y_1$ with $Y_2$, $Q_1$ with $Q_4$, $Y_3$ with $Y_4$, $Q_3$ with $Q_6$ and so on. This is represented in Fig. \ref{Multiverse}.

\begin{figure}[h]
  \centering
  \includegraphics[width=.8\linewidth]{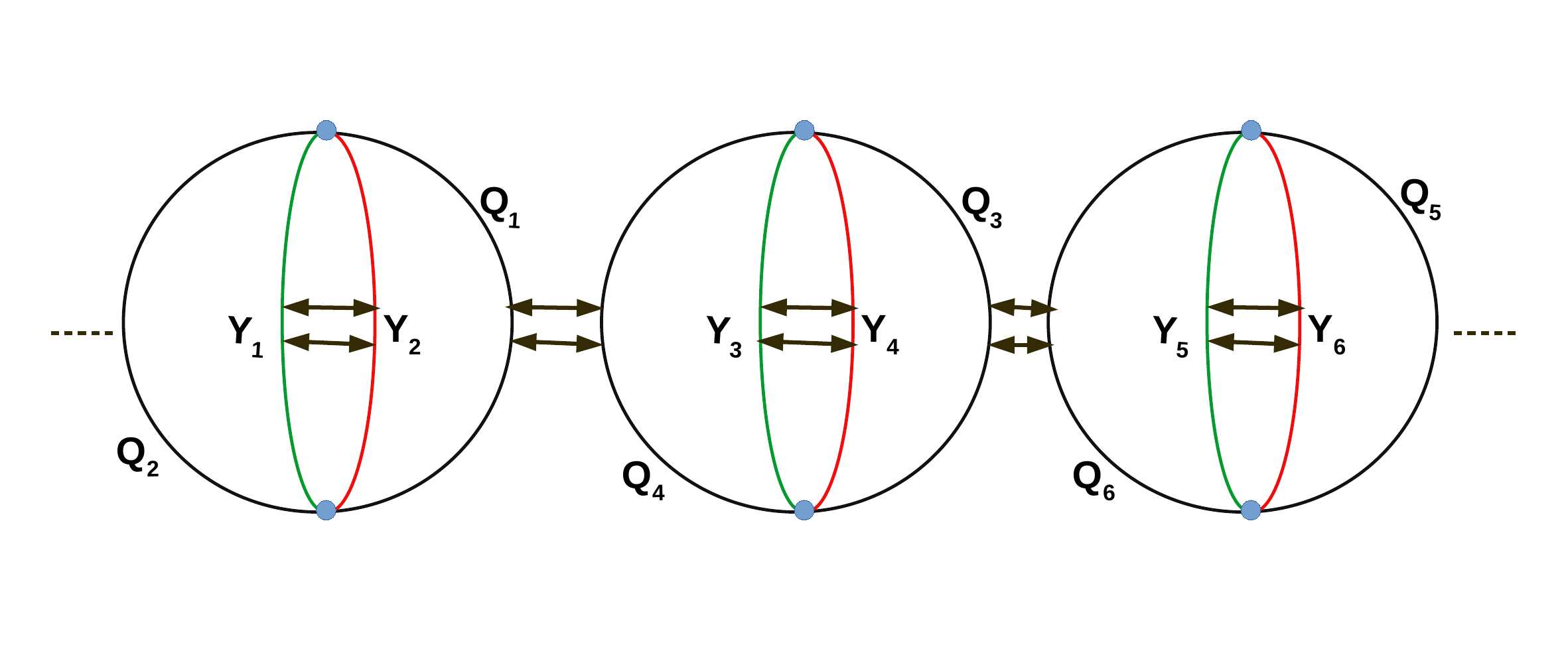}
\caption{Communicating universes in de-Sitter wedge holography. ``.....'' in the figure represents other copies of wedge holography.}
\label{Multiverse}
\end{figure}
If we follow the above idea and join the various branes following \cite{Kawamoto:2023wzj}, then Fig. \ref{Multiverse} end up at the end of gluing as shown in Fig. \ref{Multiverse-i} where we have replaced ``sphere'' with ``box'' only for convenience so that we can draw the cartoon picture easily. The notations are, e.g., $Q_{14}=Q_1+Q_4$, $Y_{12}=Y_1+Y_2$, $Q_{36}=Q_3+Q_6$, $Y_{34}=Y_3+Y_4$ and so on; in general $Q_{ij}=Q_i+Q_j$ and $Y_{ij}=Y_i+Y_j$ where $i$ and $j$ are the labels of the branes.
\begin{figure}[h]
  \centering
  \includegraphics[width=0.7\linewidth]{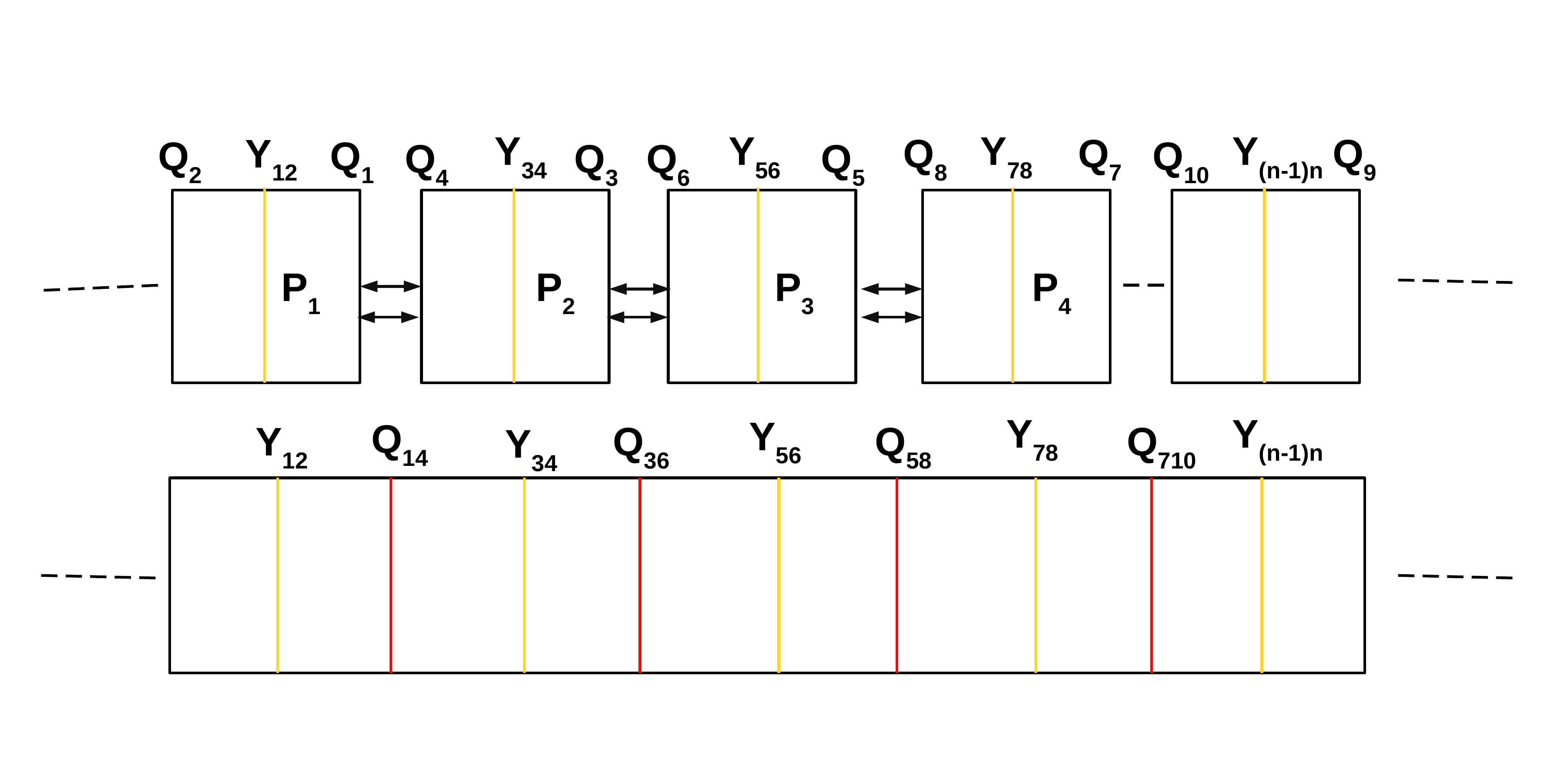}
\caption{Structure of the communicating universes in de-Sitter wedge holography. We have represented the gluing of $Y_i$'s brane ($Y_{ij}=Y_i+Y_j$) in yellow color whereas the gluing of Randall-Sundrum branes, $Q_i$'s ($Q_{ij}=Q_i+Q_j$) by red color.}
\label{Multiverse-i}
\end{figure}
Let us understand the physical meaning of Fig. \ref{Multiverse} and \ref{Multiverse-i}. In the beginning, we take two copies of double holography in DS/dS correspondence \cite{Geng:2021wcq}, and we obtain the geometrical structure as the one copy of wedge holography as shown in Fig. \ref{Multiverse} with Randall-Sundrum branes $Q_{1,2}$ and UV branes $Y_{1,2}$. When we join $Y_1$ and $Y_2$ using \cite{Kawamoto:2023wzj}, then we obtain the first box as shown in the upper part of Fig. \ref{Multiverse-i} with Randall-Sundrum branes $Q_{1}$ and $Q_{2}$ and one UV brane $Y_{12}$. This describes a single de-Sitter space with cut-off by the Randall-Sundrum branes $Q_{1}$ and $Q_{2}$. We can treat this kind of box as a building block (say $P_1$) and then join many copies of this building block as shown in the upper part of Fig. \ref{Multiverse-i}, e.g., $P_1$ with $P_2$, $P_2$ with $P_3$, $P_3$ with $P_4$ and so on. $P_i$'s are like a single de-Sitter space with two Randall-Sundrum branes. As long as $ P_i$s are separated, this model will provide the structure of parallel universes. Whenever we join all the $P_i$'s with each other, then we obtain the lower part of Fig. \ref{Multiverse-i}, which describes a single de-Sitter space with two Randall-Sundrum branes at the end of de-Sitter space. Hence, the existence of a parallel universe in extended static patch holography depends upon whether we join all copies of extended static patch holography or not.\par 

 For three copies, we have Randall-Sundrum branes as $Q_{1,2,3,4,5,6}$ and UV branes as $Y_{1,2,3,4,5,6}$  and this is described in Fig. \ref{Multiverse-i-n=3}.
\begin{figure}[h]
  \centering
  \includegraphics[width=0.5\linewidth]{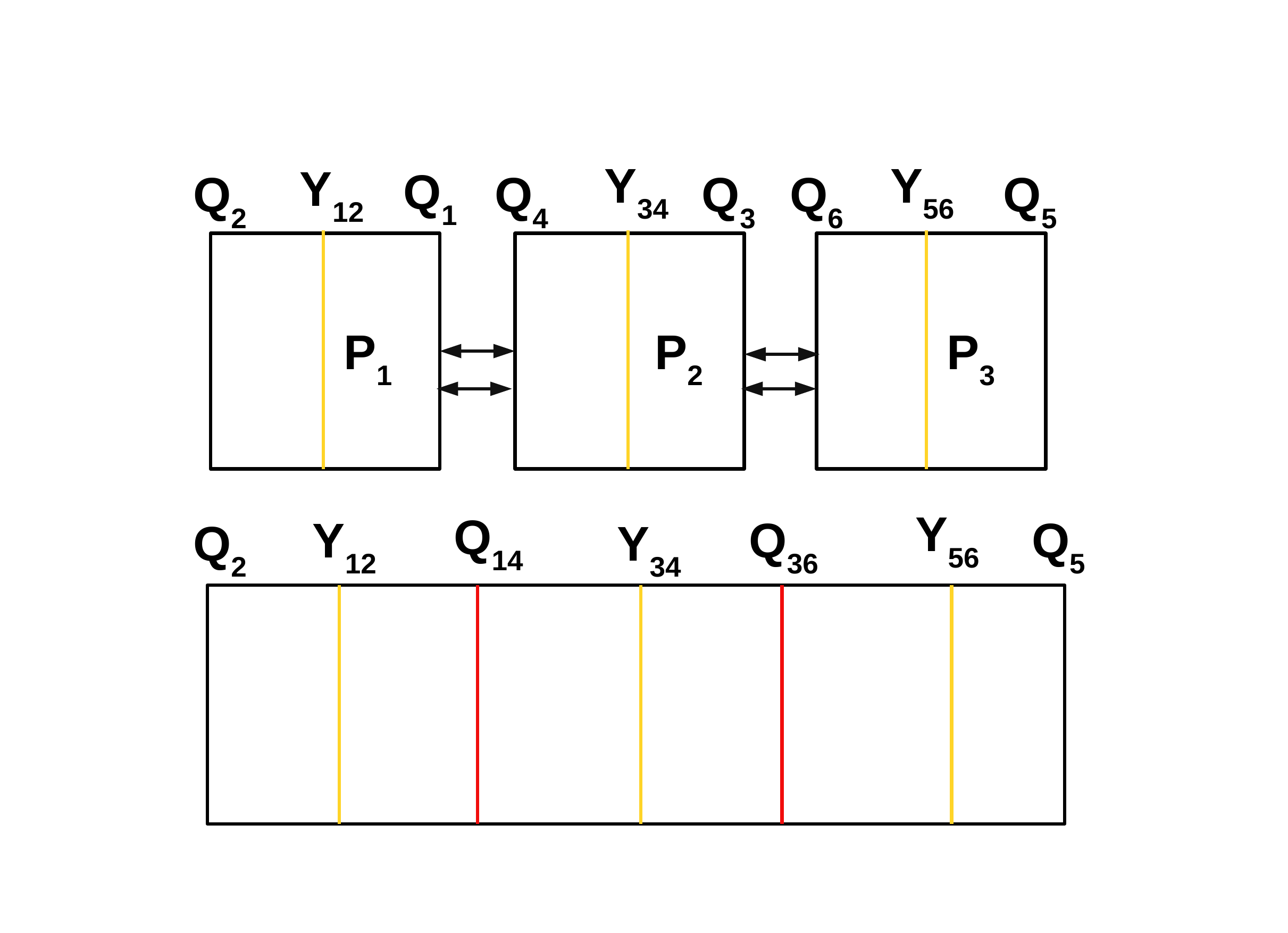}
\caption{Structure of the communicating universes in de-Sitter wedge holography for $n=3$.}
\label{Multiverse-i-n=3}
\end{figure}

\section{Communicating Multiverses in de-Sitter wedge holography}
\label{PM-dSWH}
This section discusses the existence of multiverse and communicating multiverses in de-Sitter wedge holography via subsections \ref{Multiverse-sec} and \ref{PM-subsec}.
\subsection{Multiverse}
\label{Multiverse-sec}
We can construct a multiverse for a single de-Sitter space as shown in Fig. \ref{Multiverse-single-dS}.
\begin{figure}[h]
  \centering
  \includegraphics[width=.5\linewidth]{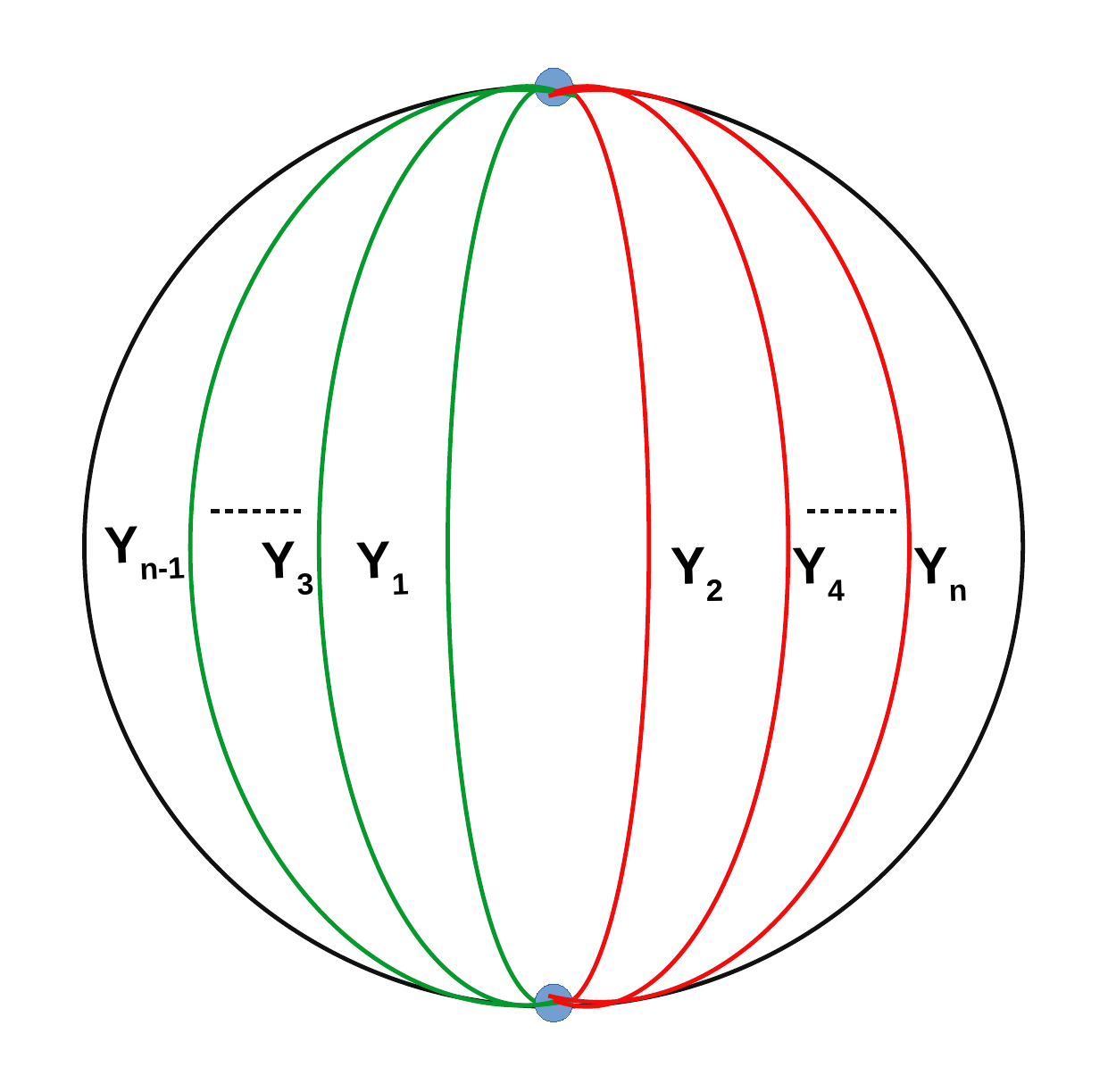}
\caption{Multiverse in de-Sitter wedge holography. $Y_{1,2,...,n-1,n}$ are the constant $\omega$ slices in a single de-Sitter space.}
\label{Multiverse-single-dS}
\end{figure}
We obtain the Multiverse model in a single de-Sitter space by taking the constant $\omega$ slices of a single de-Sitter bulk (\ref{bulk-metric-dS}). As we discussed earlier, we have Einstein gravity with positive cosmological constant on constant $\omega$ slices, and hence we have a Multiverse made up of ``$n$'' universes with geometry on each of them as dS$_{d-1}$, and all these universes are embedded in a single de-Sitter bulk with geometry dS$_d$. This kind of model for the AdS bulk was discussed in \cite{Yadav:2023qfg}. Here, we are able to construct the Multiverse model in de-Sitter space which is more realistic as we are living in a universe with positive cosmological constant. One can check that the bulk metric (\ref{bulk-metric-dS}) satisfies the Neumann boundary condition (\ref{NBC}) at $\omega=\omega_{1,2,....,n-1,n}$ provided tensions of the end-of-the-world branes should be given as follows
\begin{eqnarray}
\label{T-dS-Multiverse}
T^\beta=\frac{d-2}{R} \cot\left(\frac{\omega_{1,2,....,n-1,n}}{R} \right),
\end{eqnarray} 
where $\beta=1,2,....,n-1,n$. The interesting point about this setup is that all the universes ($Y_1$, $Y_2$,....,$Y_{n-1}$, and $Y_n$) are joined at the single defect (blue dot in Fig. \ref{Multiverse-single-dS}) and hence there is the possibility of exchange of information among each other.

\subsection{Communicating Multiverses}
\label{PM-subsec}
Given that Multiverse exists in de-Sitter wedge holography as discussed in subsection \ref{Multiverse-sec}. Now, we discuss the possibility of the existence of communicating multiverses. It is possible by joining many copies of the Multiverse as shown in Fig. \ref{Multiverse-single-dS}. 
\begin{figure}[h]
  \centering
  \includegraphics[width=1.05\linewidth]{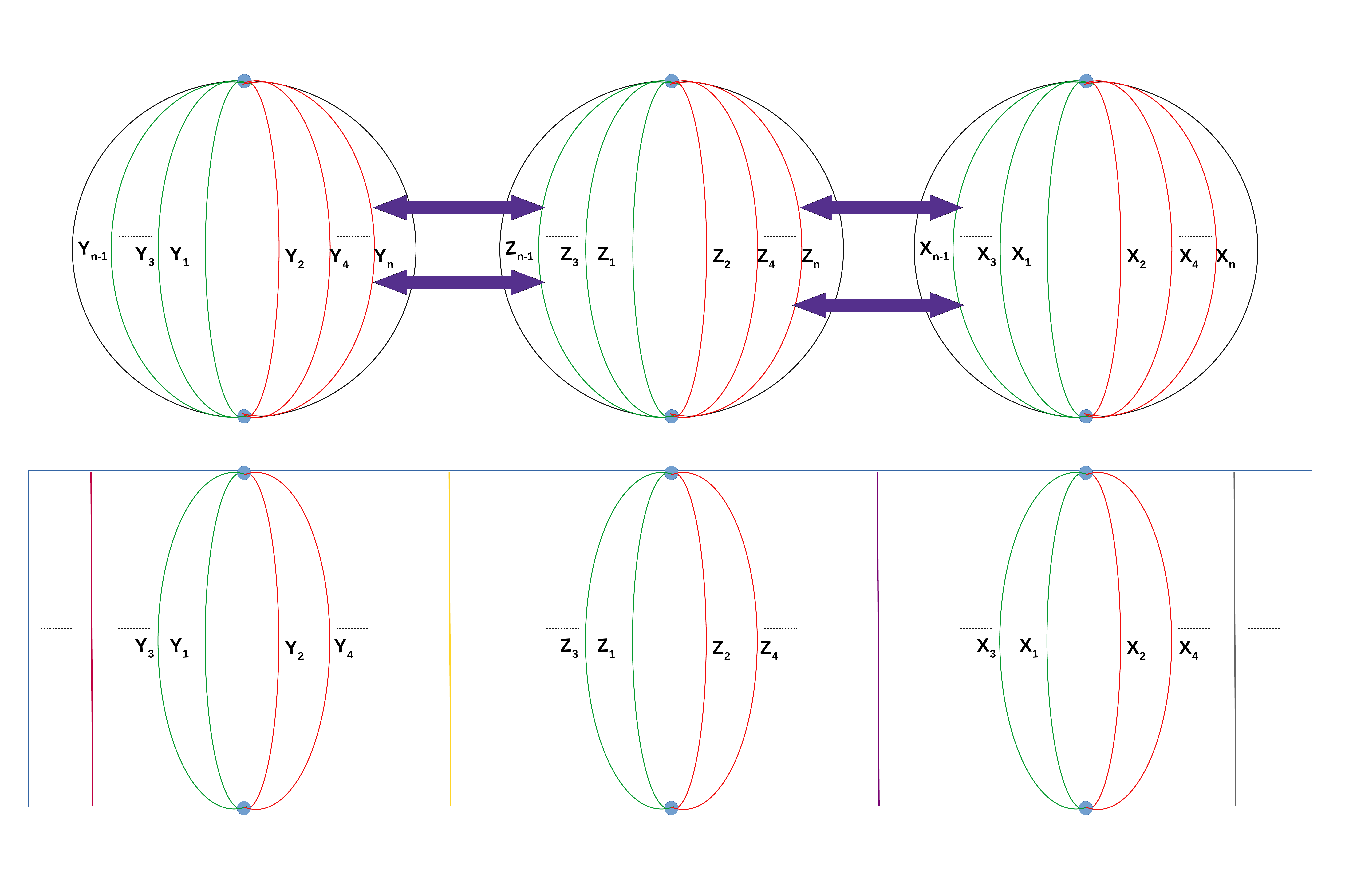}
\caption{Communicating multiverses in de-Sitter wedge holography. For simplicity, we replaced the sphere with a rectangle in the lower part of this figure.}
\label{Parallel-multiverses-dS}
\end{figure}
To do so, we join $n$th brane of one copy with $(n-1)$th brane of the second copy,  $n$th brane of the second copy with $(n-1)$th brane of the third copy, and so on; this is represented pictorially in Fig. \ref{Parallel-multiverses-dS}. More precisely, the left outermost brane of the first copy will join with the right outermost brane of the second copy; the left outermost brane of the second copy will join with the rightmost outer brane of the third copy, and so on. We have done it for three copies in Fig. \ref{Parallel-multiverses-dS}.
\par
In the process of constructing communicating multiverses, we have to lose some of the universes. For example, if we join two multiverses, then we have to lose two universes: one from the left copy($Y_n$) and one copy from the other copy ($Z_{n-1}$). These two branes will form a new brane, which is represented by a yellow line in the lower part of Fig. \ref{Parallel-multiverses-dS}. Similarly, if we join Multiverse made up of $Z_{1,2,...,n-1,n}$-branes with the Multiverse made up of $X_{1,2,....,n-1,n}$ branes, then the combination of $Z_n$ brane and $X_{n-1}$ brane will form a new brane and the same is represented by dark purple in the lower part of Fig. \ref{Parallel-multiverses-dS}. 

To have a consistent model of communicating multiverses, we have to stop the process of joining branes from the consecutive multiverses; otherwise, all left branes of one copy will join with the all right branes of the right copy, and at the end, we have a single universe as obtained in section \ref{PU-ESP}. To be more specific in the context of Fig. \ref{Parallel-multiverses-dS}, all the left branes of multiverse ``$Y$''(which consist of universes $Y_{1,2,...,n-1,n}$) will join with all the right branes of multiverse ``$Z$'' consists of universes $Z_{1,2,...,n-1,n}$; all the left branes of multiverse ``$Z$'' with universes $Z_{1,2,...,n-1,n}$ will join with all the right branes of multiverse ``$X$'' with universes $X_{1,2,...,n-1,n}$ and so on. In the end, we obtain a single de-Sitter space bounded by two end-of-the-world branes.

\section{Comments on de-Sitter holography}
\label{dS-CFT correspondance}

Our motivation to construct wedge holography in de-Sitter space is inspired by the wedge holography construction in AdS spacetime \cite{WH-1,WH-2}. The idea for this construction is as follows. Suppose we have $(d+1)$-dimensional AdS gravity in the bulk. The first step is to localize this bulk gravity on two gravitating Karch-Randall branes in $d$-dimensions. These two branes are joined at the $(d-1)$-dimensional defect. The same is described as follows:\\ 
~\\ 
{\large
\fbox{\begin{minipage}{34em}
{\it 
Classical gravity in $(d+1)$-dimensional AdS spacetime \\ \hspace{2cm}   $\updownarrow$ \tcb{\footnotesize braneworld holography in AdS spacetime \cite{KR1,KR2}}
~\\   $\equiv$ (Quantum) gravity on two $d$-dimensional Karch-Randall branes with metric AdS$_{d}$\\ \hspace{2cm} $\updownarrow$ \tcb{\footnotesize AdS$_d$/CFT$_{d-1}$ correspondence \cite{AdS-CFT}}
~\\ $\equiv$ CFT living at the $(d-1)$-dimensional defect.}\end{minipage}}}\\
~\\
Mathematically, $AdSW_{d+1}/CFT_{d-1}$ implies that 
\begin{equation}
\label{z1}
Z_{{\rm CFT}_{d-1}}=e^{- I_{{\rm AdSW}_{d+1}}},
\end{equation}  
where $Z_{{\rm CFT}_{d-1}}$ is the partition function of $CFT_{d-1}$ in large $N$ limit and  $I_{{\rm AdSW}_{d+1}}$ is the classical gravitational action in $(d+1)$ dimensional asymptotically AdS spacetime with a wedge. Also, $AdS_d/CFT_{d-1}$ states that 
\begin{equation}
\label{z2}
Z_{{\rm CFT}_{d-1}}=e^{- I_{{\rm AdS}_{d}}}.
\end{equation}  
The author in \cite{WH-1,WH-2} proved that $I_{{\rm AdSW}_{d+1}}= I_{{\rm AdS}_{d}}$ with effective Newton constant in $d$-dimensions: $\frac{1}{G_N^{(d)}}=\frac{1}{G_N^{(d+1)}}\int_0^{\rho}dx \cosh^{d-2}(x)$ where ``$\cosh(x)$'' appears in the bulk metric as given below:
\begin{equation}
\label{metric-RXM}
ds^2=g_{\mu \nu} dx^\mu dx^\nu=dx^2 +\cosh^2(x)\hat{h}_{ij}(y)dy^i dy^j,
\end{equation}
where $x^\mu = (x, y^i)$, $g_{\mu \nu}$ and $\hat{h}_{ij}$ are the metrics in $(d+1)$ dimensions and $d$ dimensions, respectively. In \cite{WH-2}, branes are located at $x=\pm \rho$. Hence (\ref{z1}) is true. In the same paper \cite{WH-2}, the author has shown that for a different warp factor ($\sinh^2(x)$) in the metric (\ref{metric-RXM}), one can obtain the de-Sitter gravity on the Karch-Randall branes for the particular form of the tensions and effective Newton constant on these branes. Here it was explicitly shown that $I_{{\rm AdSW}_{d+1}}= I_{{\rm dS}_{d}}$ with effective Newton constant: $\frac{1}{G_N^{(d)}}=\frac{1}{G_N^{(d+1)}}\int_0^{\rho}dx \sinh^{d-2}(x)$. For more discussion about this type of holography, see \cite{WH-1,WH-2}.
~\\
\par
We follow the same idea for the de-Sitter space as a bulk theory. The dictionary for the wedge holography in de-Sitter space is given as:\\
~\\
{\large
\fbox{\begin{minipage}{34em}
{\it 
Classical gravity in $d$-dimensional de-Sitter space \\ \hspace{2cm}   $\updownarrow$ \tcb{\footnotesize ``braneworld-like holography in de-Sitter space'' \cite{Karch-dS}/ ``DS/dS correspondence'' \cite{Alishahiha:2004md}}
~\\   $\equiv$ (Quantum) gravity on two $(d-1)$-dimensional EOW branes with metric dS$_{d-1}$\\ \hspace{2cm} $\updownarrow$ \tcb{\footnotesize ``dS/CFT correspondence'' \cite{Strominger:2001pn,dS-CFT-1}/ ``static patch holography'' \cite{Susskind:2021omt,Franken:2023pni}/~``DS/dS correspondence'' \cite{Alishahiha:2004md}}
~\\ $\equiv$ CFT living at the $(d-2)$-dimensional defect.}\end{minipage}}}
\vspace{5mm}

Hence, {\it classical gravity in $d$-dimensional de-Sitter space is dual to $(d-2)$-dimensional defect CFT living at the corner of the wedge, which is the signature of dS$_d$/CFT$_{d-2}$ correspondence.} The CFT dual of de-Sitter space is a non-unitray theory which is Euclidean CFT (ECFT)  if we apply \cite{Strominger:2001pn,dS-CFT-1}\footnote{Good observables in this type of de-Sitter holography are the meta-observables \cite{Witten:2001kn}.} which states that ``$\Psi[g]=Z[g]$'' where $\Psi[g]$ and $Z[g]$ are the wavefunction of the universe \cite{Hartle:1983ai} and partition function of dual CFT for a given metric $g$. We can evaluate $\Psi[g]$ as $\Psi[g_{ij}] \propto e^{\iota I[g_{\mu \nu}]}$ where $g_{ij}$ is defined on the three-dimensional spatial slice of four-dimensional no boundary geometry with metric $g_{\mu \nu}$ \cite{Halliwell:2018ejl}. Mathematically, $dSW_{d}/CFT_{d-2}$ implies that 
\begin{equation}
\label{z1-dS}
Z_{{\rm CFT}_{d-2}}[\hat{h}_{ij}]=e^{\iota I_{{\rm dSW}_{d}}[g_{MN}]}.
\end{equation}  
According to dS/CFT duality \cite{Strominger:2001pn,dS-CFT-1}, 
\begin{equation}
\label{z2-dS}
Z_{{\rm CFT}_{d-2}}[\hat{h}_{ij}]=e^{\iota I_{{\rm dS}_{d-1}}[\hat{g}_{\mu \nu}]},
\end{equation} 
where $Z_{{\rm CFT}_{d-2}}[\hat{h}_{ij}]=\Psi_{\rm dS}^{(d-2)}[\hat{h}_{ij}]=e^{\iota I_{{\rm dS}_{d-1}}[\hat{g}_{\mu \nu}]}$ with $\Psi_{\rm dS}$ being the wave function of the universe for de-Sitter space, and the same can be evaluated using the no boundary proposal \cite{Halliwell:2018ejl}. The metric with different indices appearing in (\ref{z1-dS}) and (\ref{z2-dS}) can be read-off from the following equation
\begin{eqnarray}
\label{bulk-metric-nb}
& & \hskip -1.05in ds_{\rm global \ de-Sitter}^2=g_{MN}dx^M dx^N= d \omega^2+ \sin^2\left(\omega/R\right) \hat{g}_{\mu \nu} dx^\mu dx^\nu \nonumber\\
& & = d \omega^2+ \sin^2\left(\omega/R\right) \left(-d\tau^2+ \cosh^2\left(\tau\right) \hat{h}_{ij} dx^i dx^j \right)\nonumber\\
& &  =d \omega^2+ \sin^2\left(\omega/R\right) \left[-d \tau^2+\cosh^2\left(\tau\right) d \Omega_{d-2}^2 \right].
\end{eqnarray}

 From (\ref{z1-dS}) and (\ref{z2-dS}), we can see that if we can prove that $I_{{\rm dSW}_{d}}[g_{MN}]=I_{{\rm dS}_{d-1}}[\hat{g}_{\mu \nu}]$ then the codimension two holography (\ref{z1-dS}) exists in de-Sitter space. It has been shown in \cite{Geng:2019bnn} that effective Netwon constant in $(d-1)$-dimensions and Newton constant in $d$-dimensions are related via $\frac{1}{G_N^{(d-1)}}=\frac{1}{G_N^{(d)}}\int d\omega e^{(d-3) A(\omega)}$ for the general warp factor appearing in the metric of the form $ds^2_d=d\omega^2+e^{2 A(\omega)}ds^2_{d-1}$. In \cite{Karch-dS}, it has been shown that we can localize $d$-dimensional de-Sitter gravity on $(d-1)$-dimensional brane. This justifies the localization of higher dimensional de-Sitter gravity in lower dimensions similar to the AdS case \cite{KR1,KR2} or $I_{{\rm dSW}_{d}}[g_{MN}]=I_{{\rm dS}_{d-1}}[\hat{g}_{\mu \nu}]$ with effective Newton constant in $(d-1)$-dimensions $\left(\frac{1}{G_N^{(d-1)}}=\frac{1}{G_N^{(d)}}\int d\omega e^{(d-3) A(\omega)}\right)$ where for (\ref{bulk-metric-nb}), $e^{2 A(\omega)}=\sin^2\left(\omega/R\right)$. 

Therefore, to construct the wedge holography in de-Sitter space and see the existence of $dS_d/CFT_{d-2}$ correspondence, we combined the results from the braneworld-like holography in de-Sitter space where we localized higher dimensional gravity in de-Sitter space ($dS_d$) to codimension one geometry in de-Sitter space ($dS_{d-1}$) \cite{Karch-dS} and then the localized gravity ($dS_{d-1}$) is dual to CFT living at the future boundary of de-Sitter space in the global coordinate because of dS/CFT correspondence \cite{Strominger:2001pn,dS-CFT-1}. Hence, the nature of CFT is the same as \cite{Strominger:2001pn,dS-CFT-1}; the only difference is that we have codimension two holography: the gravity living in the wedge formed by two end-of-the-world branes is dual to Euclidean CFT living at a defect (see Fig. \ref{WH-single-dS}). The metric on end-of-the-world branes located at $\omega=\omega_{1,2}$ in this case is: 
\begin{eqnarray}
\label{EOW-metric-global-dS}
& & ds^2 = -d \tau^2+\cosh^2\left(\tau\right) d \Omega_{d-2}^2.
\end{eqnarray}
Hence, CFT in the above discussion will be similar to \cite{Strominger:2001pn,dS-CFT-1} living at the future/past boundary of (\ref{EOW-metric-global-dS}). There is criticism of this type of de-Sitter holography that the existence of Poincare recurrences at extremely late time scales, implied by the finiteness of de Sitter entropy, prevents the existence of local observables in the infinite time limit \cite{Dyson:2002nt}. This criticism may not be resolved as the the CFT description in this case is based on \cite{Strominger:2001pn,dS-CFT-1}.\\
\par

 As discussed in section \ref{WHMD}, localized gravity can also have the structure of de-Sitter space static patch metric. This indicates that here one has to use the notion of static patch holography \cite{Susskind:2021omt,Franken:2023pni} to discuss the CFT living at the corner of wedge in Fig. \ref{WH-single-dS}. End-of-the-world branes located at $\omega=\omega_{1,2}$ have the following metric in this case\footnote{We set $l=1$ in the horizon.}:
\begin{eqnarray}
\label{EOW-metric-static-patch}
& & ds^2 =-(1-r^2)dt^2+\frac{dr^2}{1-r^2}+r^2 d \Omega_{d-3}^2.
\end{eqnarray}

An alternative way to understand this duality is from the perspective of DS/dS correspondence \cite{Alishahiha:2004md}, which implies that classical gravity in $d$ dimensional de-Sitter space is dual to two $(d-1)$ dimensional CFTs coupled to each other and $(d-1)$ dimensional de-Sitter gravity at $\omega=\pi R/2$ slice. DS/dS duality has been explored in more detail in the papers \cite{Dong:2018cuv,Gorbenko:2018oov,Lewkowycz:2019xse,Geng:2019bnn,Geng:2019zsx,Geng:2019ruz,Geng:2020kxh}. In these papers; authors studied the DS/dS correspondence from quantum information theoretic point of view where they have calculated entanglement entropy, Renyi entropy, complexity, Entanglement of Purification (EoP) in the absence as well as presence of $T\overline{T}$ deformation. To be specific, authors in \cite{Geng:2019bnn} have discussed swampland bound on the entanglement entropy which says that $\frac{S_{D,static}}{S_{dS}}\leq 1$ and $\frac{S_{D,global}}{S_{dS}}\leq 2$ where $S_{D,static}$, $S_{D,global}$, and $S_{dS}$ are the entanglement entropy of de-Sitter space in static patch, global coordinate and de-Sitter entropy respectively. Since in DS/dS correspondence, we have two CFT sectors (say ``L'' and ``R''), and they are coupled to each other at the UV slice ($\omega=\frac{\pi R}{2}$); entanglement in this holographic theory is described as entanglement between ``L'' and ``R'' sectors, and if one wants to compute the entanglement entropy then one has to trace out one sector and then one can define von Neumann entropy and this has been precisely verified in \cite{Dong:2018cuv} from the calculations in field theory as well bulk side from holographic point of view; see also \cite{Lewkowycz:2019xse} for the precise agreement between deformed-CFT and gravity calculations. In the presence of $T\overline{T}$ deformation, entanglement entropy was calculated in \cite{Gorbenko:2018oov}. Complexity and Entanglement of Purification (EoP) were studied in \cite{Geng:2019ruz}, where the author has also argued that theory living at the central slice is non-integrable. See \cite{Geng:2020kxh} for the study of shock wave and OTOC in DS/dS correspondence. From these studies and progress in DS/dS correspondence, we are able to learn about de-Sitter holography to some extent. 

The DS/dS correspondence gives a kind of effective field theory description for the de-Sitter space as mentioned explicitily in \cite{Alishahiha:2004md} that dimensional reduction from $d$ to $(d-1)$-dimensions results in effective field theory with a finite $(d-1)$-dimensional Planck mass, i.e., dynamical gravity is included in $(d-1)$-dimensional theory and also the swampland distance conjecture \cite{Ooguri:2018wrx} is valid in DS/dS correspondence \cite{Geng:2019zsx} which says that ``there is breaking of effective field theory description if moduli(scalar field) moves a large distance in field (moduli) space''. The distance conjecture restricts the size of the Hilbert space. For more discussion on this issue, see \cite{Geng:2019zsx,Ooguri:2018wrx}\footnote{We thank Hao Geng for discussion on this topic.}. However, \cite{Alishahiha:2004md} describes the energy regime $0<E \leq \frac{1}{R}$; for further study on DS/dS correspondence in the energy regime $ \frac{1}{R}<E < M_d$, see \cite{Alishahiha:2005dj} where $M_d$ is the Planck mass scale for bulk $d$-dimensional de-Sitter gravity. In this paper, we restrict ourselves to the regime $0<E \leq \frac{1}{R}$. Further, DS/dS correspondance is valid on short timescales in comparison to Poincare recurrence times and the discreteness of energy spectrum due to the finiteness of de-Sitter entropy becomes important at the Poincare recurrence times \cite{Dyson:2002nt,Goheer:2002vf}. So in the DS/dS correspondence, we may overcome the issue imposed by \cite{Dyson:2002nt,Goheer:2002vf}, see also \cite{Klemm:2004mb,Aguilar-Gutierrez:2023hls,Maldacena:2024uhs} where other methods have been proposed to resolve the criticism of \cite{Dyson:2002nt,Goheer:2002vf}.

See also an interesting paper \cite{Shaghoulian:2021cef}(also known as bilayer proposal, which is the modification of monolayer proposal \cite{Susskind:2021dfc,Susskind:2021esx}) where the author has discussed the generalized entropy formula in de-Sitter space. This supports EFT description\footnote{See also \cite{Brahma:2020tak,Bernardo:2021rul} for EFT description of de-Sitter space.} in de-Sitter space in a similar manner when the island formula was discussed in \cite{Almheiri:2019hni,Almheiri:2020cfm} where AdS gravity couples with CFT living on end-of-the-world brane. The discussion in \cite{Shaghoulian:2021cef} is based on static patch holography. The author has explicitly wrote down the formula for the generalized entropy in de-Sitter space by coupling Einstein gravity with positive cosmological constant with CFT$_d$ as
\begin{eqnarray}
\label{gen-entropy-dS}
& & S_{\rm gen}=\frac{{\rm Area}(S^{d-2})r_1^{d-2}}{4 G_N^{(d)}}+\hat{S}_{matter}(\tilde{I}),
\end{eqnarray}
where $\tilde{I}$ is defined as belt-like region on the sphere $S^{d-1}$ at time $t=0$ with $r_1$ and $r_2$ as left and right endpoints for the following metric:
\begin{equation}
\label{metric-SP+ESP}
ds^2=-(1-r^2)dt^2+\frac{dr^2}{(1-r^2)}+r^2 d\Omega_{d-2}^2=-\cos^2\theta dt^2+d\theta^2+sin^2\theta d\Omega_{d-2}^2,
\end{equation}
where $r=\sin \theta$ with $\theta \in [0,\pi/2]$ covers one static patch, and we can see that the neighboring static patch is covered by $\theta \in [\pi/2,\pi]$. The belt-like region is defined by $r \in (r_1,r_2)$ or $\theta \in (\theta_1,\theta_2)$. For the pictorial description of the belt-like region, see Fig. {\bf 3} of \cite{Shaghoulian:2021cef}. The author has obtained, in a nice way, a transition from the Hartman-Maldacena to the island surfaces transition; see Fig. {\bf 7} of \cite{Shaghoulian:2021cef}. Overall \cite{Shaghoulian:2021cef} provides many useful insights about de-Sitter space and what we have learned about black holes. One point that we would like to point is that what the author has discussed in \cite{Shaghoulian:2021cef} has been further advanced in this paper as we have the concrete wedge holographic realization of the extended static patch (section \ref{WH-ESP-sec}). Another point is that we also have the wedge holographic realization of de-Sitter space with a global de-Sitter metric, which was not present in the literature. It would be nice to study the generalized entropy formula for this case similar to \cite{Shaghoulian:2021cef}.

The double holography has been constructed in DS/dS correspondence in \cite{Geng:2021wcq}; we also discussed in section \ref{WH-ESP-sec} that wedge holography is obtained by gluing two copies of DS/dS correspondence, see description {\bf d} in Fig. \ref{WH-ESP}. The wedge holography in the context of DS/dS correspondence is described as follows. We have $(d+1)$-dimensional de-Sitter gravity in bulk, and this bulk gravity is localized on Randall-Sundrum branes $Q_1$ and $Q_2$ in $d$-dimensions because of DS/dS correspondence \cite{Alishahiha:2004md}. We can dualize $d$-dimensional de-Sitter gravity living on Randall-Sundrum branes to $(d-1)$-dimensional CFT living at the corner of the wedge formed by Randall-Sundrum branes $Q_1$ and $Q_2$. The question is, can we again apply DS/dS correspondence \cite{Alishahiha:2004md} to dualize $d$-dimensional de-Sitter gravity on $Q_1$ and $Q_2$?  The answer is yes as discussed in \cite{Alishahiha:2005dj}. Let us now dualize the gravity living on EOW branes using the DS/dS correspondance. If we do so then in the framework of wedge holography, we obtain $DS_d/dS_{d-2}$ correspondence which gives defect CFT as appear in \cite{Alishahiha:2004md}. Since we have extended static patch metric on $Q_1$ and $Q_2$. Hence, we can also apply static patch holography to dualize de-Sitter gravity of $Q_1$ and $Q_2$. Therefore, in this case, CFT living at the wedge formed by $Q_1$ and $Q_2$ will be similar to the ones that appear in static patch holography \cite{Susskind:2021omt,Franken:2023pni}.

Based on the above discussions. We conclude that CFT appearing in the wedge holography for the de-Sitter space can be either one appearing in \cite{Strominger:2001pn,dS-CFT-1} or the ones that appear in static patch holography \cite{Susskind:2021omt,Franken:2023pni} or the one which appear in \cite{Alishahiha:2004md} depending upon what metric we are using on EOW branes. In some sense, this setup unifies holographic proposals of de-Sitter space. Here, we have codimension two holography instead of codimension one holography in the usual AdS/CFT correspondence. Hence, we expect that whatever problems appear in the aforementioned three types of de-Sitter holography may also be present in our setup. It's nice to see that DS/dS correspondance \cite{Alishahiha:2004md} gives EFT description for the de-Sitter space as well as resolves the criticism of \cite{Dyson:2002nt,Goheer:2002vf} as discussed earlier. For the time being we can say that we have $DS_d/dS_{d-2}$ correspondence which is free from the criticism. At the moment, the precise dictionary has not been obtained, but in the future, this could be done, and then we may be able to know more about de-Sitter holography. The precise dictionary meaning matching of calculations in $d$-dimensional de-Sitter gravity and $(d-2)$-dimensional defect CFT. One thing is expected that $c=c_{Q_1}+c_{Q_2}$ similar to AdS bulk spacetime \cite{WH-1} where $c$, $c_{Q_1}$, and $c_{Q_2}$ are the central charges of CFT$_{d-2}$, CFT existing at $\omega_{1,2}=(\pi R)/2$ slices in DS/dS correspondance on Randall-Sundrum branes $Q_1$ and $Q_2$. 
\par
In $DS_{d+1}/dS_d$ correspondance, entropy is given as \cite{Dong:2018cuv}:
\begin{eqnarray}
\label{S}
& & {\cal S}=\frac{R_{\rm dS}^{d-1}}{4 G_N^{(d+1)}} \frac{2 \pi^{d/2}}{\Gamma(d/2)}.
\end{eqnarray}
For $d=3$, (\ref{S}) gives
\begin{eqnarray}
\label{S-i}
& & {\cal S}=\frac{R_{\rm dS}^{2}}{4 G_N^{(4)}} \frac{2 \pi^{3/2}}{\Gamma(3/2)}.
\end{eqnarray}
For $d=2$, (\ref{S}) gives
\begin{eqnarray}
\label{S-ii}
& & {\cal S}=\frac{2 \pi R_{\rm dS}}{4 G_N^{(3)}}=\frac{\pi c}{3},
\end{eqnarray}
where central charge $c= \frac{3 R_{\rm dS}}{2 G_N^{(3)}}  \sim \frac{R_{\rm dS}}{G_N^{(3)}}$.
Equation (\ref{S-i}) implies that central charge in $DS_4/dS_3$ scales as: $c \sim \frac{R_{\rm dS}^{2}}{G_N^{(4)}}$ and for $dS_3$ on EOW branes, central charges scale as:  $c_{Q_1/Q_2} \sim  \frac{R_{\rm dS}}{G_N^{(3)}}$ from $DS_3/dS_2$ correspondance \cite{Dong:2018cuv}}. Now using $\frac{1}{G_N^{(d-1)}}=\frac{1}{G_N^{(d)}}\int d\omega e^{(d-3) A(\omega)}$ \cite{Geng:2019bnn} for $d=4$, $\frac{1}{G_N^{(3)}}=\frac{1}{G_N^{(4)}}\int d\omega e^{ A(\omega)}$  with $e^{A(\omega)}=\sin\left(\omega/R_{\rm dS}\right)$ implying $\frac{1}{G_N^{(3)}}\sim \frac{R_{\rm dS}}{G_N^{(4)}}$. Hence $c_{Q_1}+c_{Q_2} \sim  \frac{R_{\rm dS}}{G_N^{(3)}} \sim   \frac{R_{\rm dS}^{2}}{G_N^{(4)}}$ which suggests that $c \sim c_{Q_1}+c_{Q_2}$ consistent with \cite{WH-1} analogous to the wedge holography for the AdS bulk spacetime. This is one consistency check of $DS_4/dS_2$ correspondence, and it is valid in the IR sector of matter CFT on $dS_2$ \cite{Dong:2018cuv}. The central charge vanishes in the energy regime $ \frac{1}{R}<E < M_d$ \cite{Alishahiha:2005dj}. To get more insight about $dS_d/CFT_{d-2}$ correspondance for the CFT appearing in dS/CFT correspondence \cite{Strominger:2001pn,dS-CFT-1}, one has to be careful as in this type of CFT, one talks about ``time-entanglement'' \cite{Narayan:2015vda,Doi:2022iyj,Narayan:2022afv} not ``spatial entanglement'' \cite{Ryu:2006bv,Ryu:2006ef,HRT} to compute the entanglement entropy.

We made comments on de-Sitter holography, keeping in mind that this model could unexplore many issues in de-Sitter holography in the future. The aim of the paper is to construct the Multiverse model in wedge holography with de-Sitter space as a bulk theory. As a bonus, we arrived at the codimension two holography for the de-Sitter space, which was expected similar to wedge holography for the AdS spacetime bulk \cite{WH-1,WH-2}  and flat spacetime bulk \cite{FS-Holography}. Interestingly, wedge holography for the AdS bulk has been studied a lot and has helped us to study the information paradox, complexity, multiverse models, etc. Further studies of this model can also answer these types of questions for the de-Sitter space, where many things are not well understood.

\section{Discussion}
\label{conclusion}
In this paper, we have constructed the wedge holography for the de-Sitter space as a bulk theory and discussed its application to multiverse models. First, we did this for general de-Sitter background using the idea similar to AdS bulk in section \ref{WHMD} where we used the localization of de-Sitter gravity \cite{Karch-dS} on the brane. A clue to the dS$_d$/CFT$_{d-2}$ duality is provided by this: $Z_{{\rm CFT}_{d-2}}[\hat{h}_{ij}]=e^{\iota I_{{\rm dSW}_{d}}[g_{MN}]}$ where ``$g_{MN}$'' is the $d$-dimensional de-Sitter bulk metric. Depending upon the metric on EOW branes, $CFT_{d-2}$ appearing in this codimension two holography is either non-unitary CFT appearing in \cite{Strominger:2001pn,dS-CFT-1} or CFT which appears in static patch holography  \cite{Susskind:2021omt,Franken:2023pni} or the one that appears in \cite{Alishahiha:2004md}. It would be interesting to explore this duality in more detailed way similar to \cite{WH-1,WH-2} and \cite{FS-Holography}. In this case, it is possible to obtain a multiverse-like structure by taking many constant $\omega$ slices of single de-Sitter space, see Fig. \ref{Multiverse-single-dS}. We see the existence of communicating multiverses [Fig. \ref{Parallel-multiverses-dS}] by gluing many multiverses with each other following the concept of \cite{Kawamoto:2023wzj}. It must be noted that communicating multiverses exist as long as all left branes of one multiverse are not joined with the all right branes of the other multiverse. Otherwise, we obtain a single de-Sitter space bounded by two end-of-the-world branes.
\par
In section \ref{WH-ESP-sec}, we constructed wedge holography in the extended static patch. This was done using the idea of double holography in DS/dS correspondence \cite{Geng:2021wcq}. We constructed wedge holography in this case by gluing two copies of a doubly holographic setup by following \cite{Kawamoto:2023wzj}. We can have four branes in extended static patch wedge holography if we keep separating the UV branes $Y_1$ and $Y_2$. If we join them, we obtain wedge holography with two Randall-Sundrum branes $Q_1$ and $Q_2$, see Fig. \ref{WH-ESP}. It is nice to see that one can describe parallel universes from the perspective of wedge holography. This can be achieved by taking ``$n$'' copies of wedge holography and then gluing them together in parallel. Again, parallel universes exist as long as ``$n$'' copies are not completely joined, i.e., we can have {\it disconnected parallel universes}. If we join the left brane of one copy with the right brane of the other, we obtain a single de-Sitter space bounded by two Randall-Sundrum branes (Fig. \ref{Multiverse-i}).

Based on the theoretical construction of Multiverse discussed in subsection \ref{Multiverse-sec}, we can say that {\it we are living in a universe that is part of the Multiverse} because this model is constructed in de-Sitter space, which has a positive cosmological constant. Many universes exist (where different species may be living) within the same Multiverse, but we are not able to travel among other universes because of our limitations. Suppose we are living in a universe and we want to travel in a different universe, then we have to cross the defect that is common to both universes. In this paper, we have not discussed this issue, but we hope to explore it in our future work. We have made a qualitative statement on this issue in \cite{Yadav:2023qfg} for the AdS bulk spacetime.

In subsection \ref{PM-subsec}, in the process of constructing communicating multiverses. There is a possibility that we can have {\it disconnected multiverses}. In this scenario, there can't be any communication between any universe of one Multiverse and any universe of another multiverse. However, to make the communication between two multiverses, we need to have {\it connected multiverses} where one universe (left outermost) of one Multiverse will get combined with one universe(right outermost) of the other Multiverse. In this case, there could be possibility of communication between different multiverses.

It would be interesting if one could construct this kind of model from top-down holography similar to \cite{Deddo:2023oxn} where the authors discussed the existence of binary black holes from top-down triple holography\footnote{See \cite{Yadav:2022mnv,Karch:2022rvr,Uhlemann:2021nhu} for the top-down construction of double holography and \cite{Bernardo:2020lar} for the realization of de-Sitter vacua in string theory.}. So far we have discussed the existence of parallel universes, multiverses, and communicating multiverses for the identical branes, i.e., we have multiple copies of the same universe. As we discussed in \cite{Yadav:2023qfg} that issue of mismatched branes \cite{Karch:2020iit} was present even in wedge holography which prevented us to construct the wedge holographic realization of Schwarzschild de-Sitter black hole. If one could resolve the issue of mismatched branes from the perspective of \cite{Kawamoto:2023wzj} then one may obtain the Page curve of Schwarzschild de-Sitter black hole \cite{Yadav:2022jib,Goswami:2022ylc,Goswami:2023ovb}\footnote{See also \cite{Krishnan:2020fer,Choudhury:2020hil} for island in cosmology.} from wedge holography. It would be nice to include the grey-body factors \cite{Hollowood:2021lsw} and see the affect of the same on information exchange between different universes or in usual wedge holography where Hawking radiation is exchanged between gravitating branes. Further, one can use our model to discuss the information paradox in de-Sitter space by considering one brane as the source of Gibbons-Hawking radiation \cite{PhysRevD.15.2738} and another as a bath for collecting the Gibbons-Hawking radiation. 

{
\section*{Acknowledgements}
The author would like to thank the Isaac Newton Institute for Mathematical Sciences, Cambridge, for support and hospitality during the programme {\bf Bridges between holographic quantum information and quantum gravity}. I would like to thank Prof. Aron Wall for inspiring conversations during the aforementioned programme. I am extremely grateful to the Physics department at Swansea University for the wonderful hospitality during my visit, especially to Prof. Timothy J. Hollowood for hosting me and fruitful discussions. 
The idea about this work originated in the event {\bf Holography@25} at ICTP-SAIFR during the conversation with Prof. Juan Maldacena and then further discussion with Prof. Timothy J. Hollowood. I would like to thank them and all the participants, speakers, and organizers of these two events. 
I would also like to thank Prof. Tadashi Takayanagi for the helpful clarification and comment on the earlier version of the draft. I would also like to thank Prof. Alok Laddha for the useful discussions.
 This work is partially supported by a grant to CMI from the Infosys Foundation and by EPSRC grant EP/R014604/1. I would also like to thank the Physics group at CMI for providing a vibrant research environment.

}


\end{document}